\documentclass[twocolumn,prb]{revtex4}
\usepackage{epsfig}
\usepackage{dcolumn}
\usepackage{amsmath}
\begin{document}

\title{Transport in one dimensional Coulomb gases: From ion
channels to nanopores}

\author{A.~Kamenev$^1$, J.~Zhang$^1$,  A.~I.~Larkin$^{1,2}$, B.~I.~Shklovskii$^{1,2}$}
\address{ $^{1}$ Department of Physics, University of Minnesota,
Minneapolis, MN 55455, USA\\
$^{2}$ William I. Fine Theoretical Physics Institute, University
of Minnesota, Minneapolis, MN 55455, USA}


\date{\today}
\begin{abstract}
We consider a class of systems where,  due to the large mismatch
of dielectric constants, the Coulomb interaction is approximately
one--dimensional. Examples include ion channels in lipid membranes
and water filled nanopores in silicon films. Charge transport
across such systems possesses the activation behavior associated
with the large electrostatic self--energy of a charge placed
inside the channel. We show here that the activation barrier
exhibits non--trivial dependence on the salt concentration in the
surrounding water solution and on the length and radius of the
channel.
\end{abstract}

\maketitle


\maketitle

\section{Introduction}
\label{sec1}

In a number of quasi--one--dimensional systems the interactions
between  charged carriers follow (for a certain range of
distances) the {\em one dimensional} Coulomb law: $\Phi(x)\sim
|x|$, well known for parallel charged planes. Examples include:
ion channels in biological lipid
membranes~\cite{Stryer,ionchanbook}, water--filled nanopores in
membranes for desalination devices~\cite{desalination}, water
filled nanopores in silicon oxide films~\cite{Li,Storm,Timp},
free--standing silicon nanowires \cite{Lieber,Peeters} and others.
Their common feature is the presence of a quasi 1d channel with
the dielectric constant greatly exceeding that of the surrounding
3d media. As a result, the electric field is forced to stay inside
the channel, leading to the 1d interaction potential, mentioned
above. The purpose of this paper is to study the charge transport
through such systems as a function of the carrier concentration,
length, temperature, etc. We show that these systems posses very
peculiar transport properties, qualitatively different from those
found in the examples with shorter range interactions.

To be specific, we focus on the water filled channels. One example
is ion channels in lipid membranes. Such a membrane consists of a
$L=5\, \mbox {nm}$ thick hydrocarbon layer with the dielectric
constant $\kappa_2\simeq 2$, surrounded by water with the
dielectric constant $\kappa_1 \simeq 80$. Due to the large ratio
$\kappa_1/\kappa_2 \simeq 40\gg 1$ the electrostatic self--energy
of a charged ion inside the hydrocarbon layer is huge, making the
pure membrane impermeable for ions from water. The only way for
ions to cross the membrane is through the water--filled channels
formed by proteins embedded into the membrane
\cite{Stryer,ionchanbook}. Radiuses of such cylindrical channels,
$a$, vary from 0.3 to 0.8 nm (we are concerned only with the
passive channels without motors).

Another  example is a water--filled nanopore made in a silicon
film ~\cite{Li,Storm}. Such nanopore may have the radius $a \simeq
1$ nm and the length $L \simeq 20$ nm. Silicon oxidizes around the
channel, giving $\kappa_2 \simeq 4$. Thus, also in this case
$\kappa_1/\kappa_2=20 \gg 1$. Yet another  example is provided by
water--filled nanopores, say, with $L \simeq 40$ nm and $a\simeq
1.5$ nm in cellulose acetate films used for the inverse osmosis
desalination. In this example  $\kappa_2 \simeq 2$ and again
$\kappa_1/\kappa_2= 40 \gg 1$.

\begin{figure}[ht]
\begin{center}
\includegraphics[height=0.3\textheight]{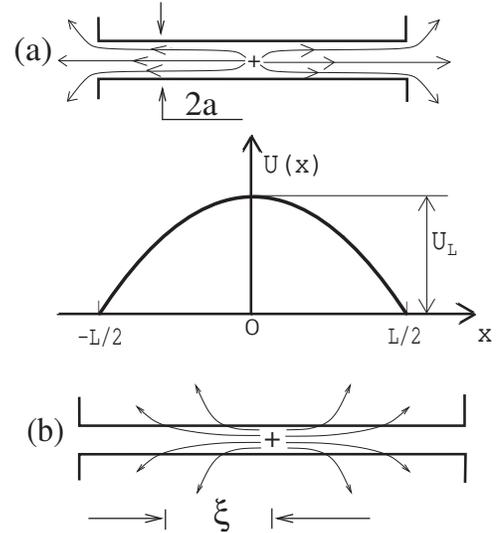}
\end{center}
\caption{ Electric field of a charge in a short (a) and a long (b)
cylindrical channel with the large dielectric constant $\kappa_1$,
in the membrane with the dielectric constant $\kappa_2<<\kappa_1$.
$L$ is the channel length, $a$ is its radius. The self--energy
barrier is shown as a function of coordinate $x$ for the case of
the short channel. $U_L$ is its maximum height. For a long channel
$\xi$ is the electric field escape length from the channel.}
\label{fig1}
\end{figure}

Keeping in mind one of these examples, let us consider a single
water filled channel in a macroscopic membrane or a film
separating two reservoirs with salty water (Fig.~\ref{fig1}a). We
assume that walls of the channel do not carry fixed charges. A
static voltage $V$, applied between the two reservoirs, drops
almost entirely in the channel due to the high conductivity of the
bulk solution. One can measure the ohmic resistance, $R$, of the
channel as a function of the temperature, $T$, concentration of
monovalent salt, $c$ (for example KCl) and parameters of the
channel $L$, $a$, $\kappa_1, \kappa_2$. The main goal of this
paper is to evaluate $R(T, c, L, a, \kappa_1, \kappa_2)$ for the
very nontrivial case: $\kappa_1 \gg \kappa_2$. This is a
completely classical ($\hbar=0$) problem.

For a low concentration of salt $c$ the charge transport is due to
the rare events, when there is a single (e.g. positive) ion inside
the channel. We assume that the radius of the ion, $b$, is smaller
than that of the channel $a$, so that the ion is totally
surrounded by water. It is easy to see that the electric field of
the ion at distances larger than $a$ is deformed because the
electric field lines avoid entering the media with the small
dielectric constant $\kappa_2$ (Fig.~\ref{fig1}). This leads to
the enhancement of the electrostatic self--energy of the ion,
which creates a barrier $U(x)$, where $x\in[-L/2,L/2]$ is the ion
coordinate  inside the channel.  This barrier is the difference
between the self--energy of the ion at the point $x$ inside the
channel and the self-energy in the bulk.  It is the maximum of the
barrier, $U(0)=U_L$, that determines the resistance of the
channel, Fig.~\ref{fig1}a.

For a relatively short channel, all field lines are confined
inside the channel until they reach its end and go out to the bulk
of the water (Fig.~\ref{fig1}a). Therefore, the electrostatic
self--energy barrier $U_L \propto L$ and does not depend on
$\kappa_2$. On the other hand, in a very long channel
(Fig.~\ref{fig1}b) the lines eventually escape to the surrounding
media with the small dielectric constant $\kappa_2$ before they
reach the exit to the bulk water. As a result, $U_L$ saturates at
some value $U_{\infty}$, which depends on the ratio
$\kappa_1/\kappa_2$, but does not depend on the length $L$.

The self--energy barrier  was studied by many
authors~\cite{Parsegian,Lev,Jordan,Fink2,Fink1,Teber}. We shall
review  their results in section \ref{sec2}. Here we calculate the
barrier for the simplest case of the short channel~\cite{Fink1}.
The electric field $E_0$ at a distance $x>a$ from a charge located
in the middle of the short channel is uniform and given by the
Gauss theorem
\begin{equation}
E_0 = {2\,e\over \kappa_1 a^{\, 2}}\, . \label{field}
\end{equation}
The energy of such field in the volume of the channel determines
the maximum height of the barrier (Fig.~\ref{fig1}a)
\begin{equation}
U_L(0) = {e^{2}L\over 2\kappa_1 a^{\, 2}}={eE_0L\over 4}\, ,
\label{short}
\end{equation}
where the zero argument  indicates that the result is valid in the
limit of vanishingly low salt concentration. For a narrow channel
the barrier, $U_{L}(0)$, can be much larger than $k_BT$, making
the resistance $R$ exponentially large:
\begin{equation}
R\sim \exp\left\{{U_L\over k_BT}\right\}\, . \label{conductance}
\end{equation}

\begin{figure}[ht]
\begin{center}
\includegraphics[height=0.065\textheight]{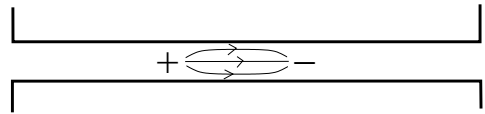}
\end{center}
\caption{Electric field of a pair of positive and negative charges
inside the channel. When charges move away from each other the
energy of the electric field between them grows linearly with the
separation. }\label{fig2}
\end{figure}

In this paper we study what happens when the concentration of the
monovalent salt, $c$, in the bulk solution increases and more ions
enter the channel. They first arrive as compact neutral pairs of
oppositely charged ions. Indeed, such  pairs can enter the channel
without the self-energy barrier. The one--dimensional character of
the electric field between the charges creates a strong
confinement between them (Fig.~\ref{fig2}).
Indeed, when a pair of oppositely charged ions separates by a
distance $x$, the uniform electric field between them, $2E_0$,
creates the confining ``string'' potential $\Phi(x)=eE_0|x|$. This
situation reminds two quarks confined in a meson. Condition
$\Phi(x)=k_BT$ defines the characteristic thermal length of the
pair:
\begin{equation}
x_T = {k_BT\over eE_0}={k_BT\kappa_1 a^{\,2}\over 2\,e^2}
={a^{\,2}\over 2\,l_{B}}\,, \label{thermallength}
\end{equation}
where $l_B \equiv  e^{2}/(\kappa_{1}k_{B}T)$ is the Bjerrum length
(for water at the room temperature $l_B=0.7\,$nm). One may think
about the pair as a one-dimensional object if $x_{T} > a$ or
$a>2l_{B}$. We first assume that this condition is fulfilled and
later return to narrower channels and show that our results are
valid (with a small modification) for the latter case, too.

At small enough concentration, $c$, (and/or temperature) the
thermal pairs do not overlap and the channel  acts as an insulator
with the exponentially large resistance, cf.
Eq.(\ref{conductance}). We study how the barrier decreases when
concentration of pairs grows. At small $c$ we calculate a
relatively small correction to the barrier, which however may be
larger than $k_BT$ and therefore leads to the exponential decrease
of $R$ with $c$. In the large $c$ limit, one may expect that the
thermal pairs overlap, screen each other and free individual
charged ions. In other words, one could expect an
insulator--to--metal (or deconfinement) transition at some
critical concentration. This does {\em not} happen, however. We
show below that, due to the peculiar nature of the 1d Coulomb
potential, the barrier proportional to the system's length
persists to any concentration of the salt, no matter how large it
is. Its magnitude, though, is exponentially suppressed at large
salt concentration, leading to the double exponential decrease of
the resistance with $c$. The absence of the phase transition is in
agreement with the earlier studies of the thermodynamics of  1d
Coulomb gas \cite{Lenard,Edwards,Schultz}.

This paper is organized as follows: Section \ref{sec2} reviews the
existing literature and formulate our main results for the effect
of salt on the channel resistance. In Section \ref{sec3} we
present  qualitative explanations of the results. Sections
\ref{sec4} and \ref{sec5} are devoted to the quantitative
derivations for the cases of short and longer channels,
correspondingly. In section \ref{sec6} we consider modifications
of the results due to the deviations from the 1d model. The latter
are essential for the very narrow channels. Section \ref{sec7}
treats the dynamics of ions inside the channels, that allows to
evaluate the pre-exponential factor in the resistance. Finally the
conclusions and examples are formulated in section \ref{sec8}.

\section{Main results}
\label{sec2}

Nontrivial physics of  the ion transport  through the narrow
channels in the lipid membranes was recognized by
Parsegian~\cite{Parsegian}. He was the first who studied the
barrier height $U_{\infty}$ in the infinitely long channel. He
wrote
\begin{equation}
U_{\infty}(0) = \frac{e^2}{\kappa_{1}a}\,  F\left({\kappa_1\over
\kappa_2}\right) \label{infinitechannelbarrier}
\end{equation}
and tabulated the function $F$. Later, numerical solutions were
obtained for a channel of a finite
length~\cite{Lev,Jordan,Nadler}. Recently~\cite{Fink2,Fink1,Teber}
approximate analytical expressions for $U_{L}(0)$ in the limits of
short channel ($L \ll \xi $) and long channel ($L \gg \xi $) were
derived along with the expression for the characteristic length of
the channel, $\xi$, separating these two regimes.

Let us review the results of Refs.~\onlinecite{Fink2,Fink1,Teber}.
We already presented their result, Eq.~(\ref{short}), for the
energy barrier of an ion in a short channel ($L \ll \xi$). For a
very long channel (Fig.~\ref{fig1}b) the electric field of the
point charge decays exponentially with $|x|$ due to the leakage of
the field lines into the media with the small dielectric constant
\begin{equation}
E(x) = E_0\,e^{-|x|/\xi}\, , \label{fielddecay}
\end{equation}
where the characteristic length $\xi$ is found as a solution of
the equation:
\begin{equation}
\xi^{2} =  a^{\,2}\left(\frac{\kappa_1}{2\kappa_2}\right) \ln
\left( {2\xi\over a} \right). \label{criticaleq}
\end{equation}
The exponential decay of Eq.~(\ref{fielddecay}) is replaced by the
true Coulomb asymptotic $E = e/(\kappa_{2 }x^{2})$, at $|x|\sim
\xi \ln(\kappa_1/\kappa_2)$, where $E\ll E_0$.

The self-energy of the point charge in the infinite channel may be
expressed through $\xi$ as:
\begin{equation}
U_{\infty}(0) = {e^{2}\xi \over \kappa_1 a^{\,2}}\, , \label{long}
\end{equation}
in  agreement with Eq.~(\ref{infinitechannelbarrier}). For the
protein channels in the lipid membranes: $\kappa_1/\kappa_2 \simeq
40$ and $\xi\simeq 6.8\, a$, see
Refs.~\onlinecite{Parsegian,Teber}.

We augment the literature results for the empty channel by showing
below that for the channels with intermediate length:
\begin{equation}
U_{L}(0)= U_{\infty}(0)\tanh\left({L\over 2\xi}\right)\, .
~\label{intermediate}
\end{equation}
This equation correctly matches Eq.~(\ref{short}) for $L \ll \xi$
and Eq.~(\ref{long}) for $L \gg \xi$. Eq.~(\ref{intermediate})
agrees with numerical calculations~\cite{Jordan} of $U_{L}(0)$
within $10\%$.

The goal of this paper is to study effects of a finite
concentration of a {\em dissociated} salt, $c$, in the bulk
solution. It is known that for a monovalent salts like KCl or NaCl
$c \simeq c_0$ at low {\em total} concentration of salt, $c_0$. At
larger $c_0>0.3$M (M means a mol per liter) the ratio $c/c_0$
decreases, but stays close to $0.8$ till the saturation limit $c_0
\sim 6$M is reached~\cite{Robinson}.

First attempts~\cite{desalination} to include screening by salt
used mean field Debye--H\"uckel or Poisson--Boltzmann
approximations. We will show below that discreteness of charge in
one-dimensional Coulomb case is so important that these
approximations fail to describe the resistance barrier.

As  explained in the Introduction, the ions easily enter the
channel in neutral compact pairs. Let us estimate the
concentration of typical thermal pairs with the arm $x_T$ inside
the channel. First, imagine a channel with $\kappa_2= \kappa_1$
and the same dimensions $L$ and $a$ as the  channel at hand. Then
the concentrations of ions in the channel is the same as outside
of it. In such a case the 1d concentration of, say, positive ions
is $n \equiv \pi a^{\,2}c$. Using $n$ we can estimate the
concentration of compact pairs in the channel as:
\begin{equation}
n_{p}=2n^{2} x_T = 2\alpha n\, ,
 \label{npairs}
\end{equation}
where
\begin{equation}
\alpha \equiv n x_T = \pi c\,a^{\,2}x_T = {\pi c\,a^{\,4}\over
2\,l_B}\,  \label{alpha}
\end{equation}
is the dimensionless concentration of dissociated salt. In
Eq.~(\ref{npairs}) at $\alpha<<1$ the factor $2\alpha$ may be
understood as the probability to find a negative ion at the
distance $x_T$ from the positive one.

In the channel with $\kappa_1\gg \kappa_2$ we find at small $c$
the actual concentration of single ions is much smaller than $n$,
so that $n$ is a purely formal construction in this case. However,
the concentration of pairs $n_p$ with the arm less than $x_T$ is
still given by Eq.~(\ref{npairs}), because these pairs enter the
channel practically without a barrier. As a result, the condition
of the small overlap of pairs may be written as $n_{p}
x_{T}=2\alpha^2 \ll 1$ or just $\alpha \ll 1 $. Thus, $\alpha$ is
the main dimensionless parameter of our theory, which describes to
what extent the pairs overlap.

\begin{figure}[ht]
\begin{center}
\includegraphics[height=0.18\textheight]{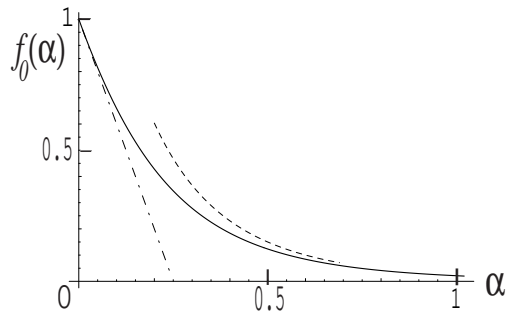}
\end{center}
\caption{ The function $f_{0}(\alpha)=U_L(\alpha)/U_L(0)$ for a
short channel. Its asymptotic limits  valid for small $\alpha$
(Eq.~(\ref{weakdop})) and large $\alpha$ (Eq.~(\ref{heavydop}))
are shown by dash-dotted and dashed lines. } \label{fig3}
\end{figure}

Now we formulate our results starting from a short channel, where
electric field lines do not escape from the channel, $L<\xi$. At
$\alpha \ll 1 $ all ions form compact pairs and the effect of
interactions between the pairs is small. Nevertheless, there is a
linear in $\alpha\ll 1$ correction to the activation barrier:
\begin{equation}
U_L(\alpha) = (1 - 4\alpha)\,U_L(0)\,, \label{weakdop}
\end{equation}
where $U_L(0)$ is given by Eq.~(\ref{short}). Although the
relative correction is small, it may have a profound influence on
the resistance of the channel. Indeed, since $4\alpha
U_L(0)/(k_BT)=nL$, according to Eq.~(\ref{conductance}):
\begin{equation}
\label{conductshort} R(\alpha\ll 1) \sim R(0) \exp\{-nL\},
\end{equation}
where $nL = c \pi a^{2}L$ is the number of ions in the volume of
the channel in the bulk solution ({\em not} the actual number of
ions in the channel). At $nL \gg 1$, the resistance is
exponentially decreased in comparison with the zero concentration
limit.

In the large $c$ limit, when $\alpha > 1 $, one may expect that
the overlap of thermal pairs leads to the deconfinement transition
at some critical concentration,  $\alpha_c \simeq 1$. We show that
such transition does {\em not} occur. Instead, the barrier
proportional to the channel's length exists at any concentration
of the salt. Its magnitude, however, is exponentially suppressed
at high salt concentration, i.e. for $\alpha \gg 1$:
\begin{equation}
U_L(\alpha) =  2^{\,7} \pi^{-1/2} \alpha^{3/4} \exp\left\{-
8\sqrt{\alpha}\right\}U_{L}(0)\,. \label{heavydop}
\end{equation}
Thus the channel is always an insulator (one may call it a {\em
super}insulator, since $\ln R \propto L$, rather than $R\propto L$
as is the case for Ohmic resistors). The $\alpha$ dependence in
Eq.~(\ref{heavydop}) is similar to that obtained in
Ref.~\cite{Schultz} for the inverse effective dielectric constant.

We plot numerically computed function $f_{0}(\alpha)\equiv
U_L(\alpha)/U_L(0)$ along with its two asymptotic
Eqs.~(\ref{weakdop}) and (\ref{heavydop}) in Fig.~\ref{fig3}. It
is  easy to verify that $f_{0}(\alpha)$ can be fitted by
$f_{0}(\alpha)\approx \exp(-4\alpha)$ (not shown) within the
accuracy of $10\%$ in the range $0<\alpha <0.94$, where
$f_{0}(\alpha)$ decreases almost 50 times.

We turn now to the results for longer channels. The short channel
results, outlined above, are valid for $L\leq L_c$, where
$L_c=L_c(\alpha)$ is a certain threshold length. At $\alpha \ll 1$
we show that $L_c=\pi\sqrt{2\alpha}\,\xi$, so the correction term
in Eq.~(\ref{weakdop}) is valid for smaller channel's length. At
$L\gg L_c$ the barrier saturates at the value
\begin{equation}
U_{\infty}(\alpha) = (1 - 4\alpha^{2}\ln(1/\alpha))\,
U_{\infty}(0)\, , \label{weakdopalpha}
\end{equation}
where $U_{\infty}(0)$ is given by Eqs.~(\ref{criticaleq}),
(\ref{long}).

\begin{figure}[ht]
\begin{center}
\includegraphics[height=0.18\textheight]{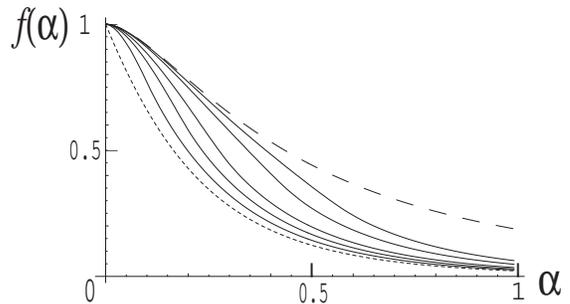}
\end{center}
\caption{ Effect of screening by salt for channels of different
lengths. Ratios $U_L(\alpha)/U_L(0)$ as functions of $\alpha$ are
plotted by the dashed line for the infinite  channel
($f_{\infty}(\alpha)$) and by the dotted line for the short
channel ($f_{0}(\alpha)$). Full lines represent the channels with
(from top to bottom) $L/\xi = 4, 3, 2, 1.5, 1$. } \label{fig4}
\end{figure}

In the $\alpha \gg 1$ limit, we obtain the exponentially large
threshold length $L_c(\alpha)\sim
\xi\exp\left\{4\sqrt{\alpha}\right\}$. Consequently,
Eq.~(\ref{heavydop}) is only valid  as long as $L \leq
L_c(\alpha)$. For even longer channels, $L\gg L_c(\alpha)$, we
obtain:
\begin{equation}
U_{\infty}(\alpha) =  2^{\,5} \pi^{-5/4} \alpha^{3/8} \exp
\left\{- 4\sqrt{\alpha}\right\} U_{\infty}(0).
\label{heavydopalpha}
\end{equation}
Fig.~\ref{fig4} shows the ratio $f_\infty(\alpha)\equiv
U_\infty(\alpha)/U_\infty(0)$ in the infinite  channel along with
that in the  short one. We also calculated $f_{L}(\alpha)\equiv
U_{L}(\alpha)/U_{L}(0)$ for several different ratios of $L/\xi$
and plotted these results in Fig.~\ref{fig4}.

For the narrow channels, $a\lesssim l_B$, all the results,
mentioned above, are still valid, provided that the dimensionless
concentration $\alpha$ is substituted by the renormalized
concentration $\alpha_{eff}$. The latter is defined and evaluated
in section \ref{sec6}.

Above we have discussed the results pertinent to the exponential
activation  term in the linear resistance. The pre--exponential
factor is discussed in section \ref{sec7}. We have also considered
the exponential suppression  of the differential resistance at
large applied voltage, $V$. For the short channel we obtain the
suppression of the activation barrier:
\begin{equation}
R(V)\sim \exp\left\{\frac{U_L(\alpha)-eV/2}{k_BT}\right\}\ ,
\label{diffconductance}
\end{equation}
where the applied voltage $eV \leq 2U_{L}(\alpha)$. For a long
channel where $L \gg L_c(\alpha)$, Eq.~(\ref{diffconductance})
(with $U_L \to U_\infty$) is valid as long as $ eV<
e^2/(\kappa_{2}L)$. In the opposite limit, $ e^2/(\kappa_{2}L) <e
V \leq 2U_{\infty}(\alpha)$, the long--range nature of the 3d
Coulomb interaction through the dielectric media with $\kappa_2$
is important. It leads to the replacement of
Eq.~(\ref{diffconductance}) by the Frenkel-Poole
law~\cite{Frenkel}
\begin{equation}
R(V)\sim\exp\left\{\frac{U_\infty(\alpha)-(e^{3}V/L
\kappa_{2})^{1/2}}{k_BT}\right\}\ . \label{Frenkel}
\end{equation}
In the next section we present simple qualitative derivations of
the above results.

\section{Qualitative consideration}
\label{sec3}

\subsection{Concept of the boundary charges}
\label{sec31}

Consider a relatively short channel where  all field lines stay
inside the channel (Fig.~\ref{fig1}a). If the average distance
between the ions is larger than $a$ each ion may be characterized
by a single coordinate $x\in [-L/2,L/2]$ along the axis of the
channel.  We deal, thus, with the 1d plasma interacting through
the one-dimensional Coulomb potential. Each ion is equivalent to a
charged plane perpendicular to the $x$--axis with the charge
density $\pm E_0/2\pi$. As  discussed in the Introduction, a
single negative ion in the middle of a channel has the
self--energy $U_L(0)$. Half of its electric field exits to the
right and half to the left. One may say that there are image
boundary charges $q=1/2$ and $q'=1/2$ on the left and right
boundaries of the channel correspondingly (hereafter all image
charges are measured in units of $e$). One can imagine that these
charges are provided by the well conducting bulk plasma in the
reservoirs.

The concept of the boundary charges, which are not supposed to be
integer is central for the present work. Let us demonstrate the
usefulness of the boundary charges by a simple example of the
ion's energy as function of its coordinate $x$. If the boundary
charge at $x=-L/2$ is $0< q <1$, then the boundary  charge at $x =
L/2$ is $q'=1-q$. Electric fields to the left and right of the
test charge is $2qE_{0}$ and $2(1-q)E_{0}$, so that the energy of
electric field $U(x;q)=(\kappa_1/2)E_{0}^{2}a^{2}[q^{2}(x+L/2)
+(1-q)^{2}(L/2-x)]$. Optimizing this energy with respect to $q$,
we find $q = 1/2 - x/L$ and for the energy as a function $x$ we
arrive at
\begin{equation}
U(x)\!=\!{\kappa_1E_{0}^{2}a^{\,2}\over 2}\left({L\over 4}
-{x^2\over L}\right)\!=\! \left(1-\left({2x\over
L}\right)^2\right)\, U_L(0)\,. \label{parabolic}
\end{equation}
This gives  the parabolic barrier with the maximum value $U_L(0)$
in the middle of the channel (see Fig.~\ref{fig1}a). The maximal
barrier corresponds to $q = q'= 1/2$.

In a similar manner  one may consider a pair of a negative ion
with the coordinate $x_1$ and a positive ion with the coordinate
$x_2$ inside the channel. They induce two boundary charges $q$ and
$q'=-q$. Writing the interaction energy of these four charges, one
finds that it reaches its optimal  value, given by
$U(x_1,x_2)=4(|x_{12}|/L-x_{12}^2/L^2)\,U_L(0)$, at $q=x_{12}/L$,
where $x_{12}\equiv x_1-x_2$. At small separation, $|x_{12}| \ll
L$ the ions attract each other with the string potential
$eE_0|x_1-x_2|$ so that typical distance between them is $x_T$.
Such a pair can contribute to the charge transport  only if the
positive and negative ions separate and move to the opposite ends
of the channel. Remarkably, the energy barrier they have to
overcome is given by $U_L(0)$. It is {\em exactly} the same as for
the single ion. The maximum is reached when $|x_{12}|=L/2$, while
$q=-q'= 1/2$.

\begin{figure}[ht]
\begin{center}
\includegraphics[height=0.0635\textheight]{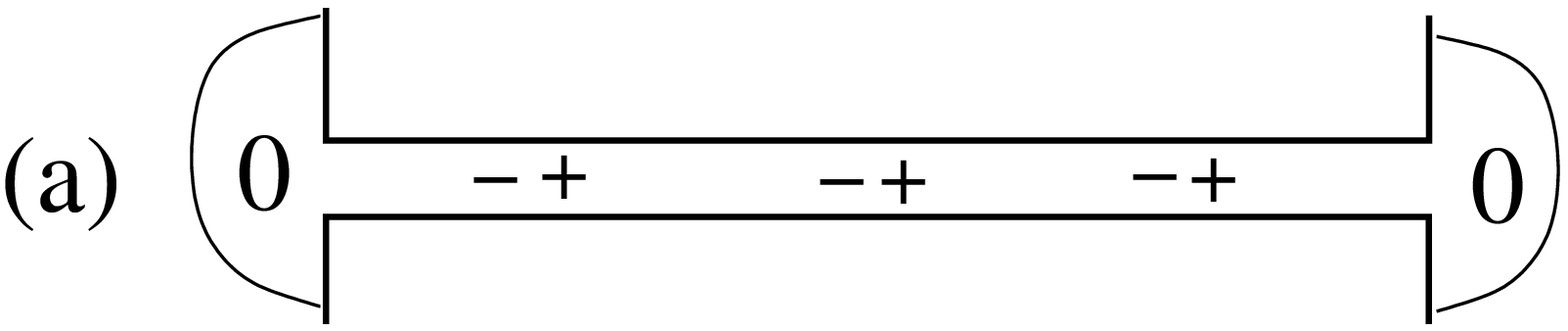} \hfill
\includegraphics[height=0.065\textheight]{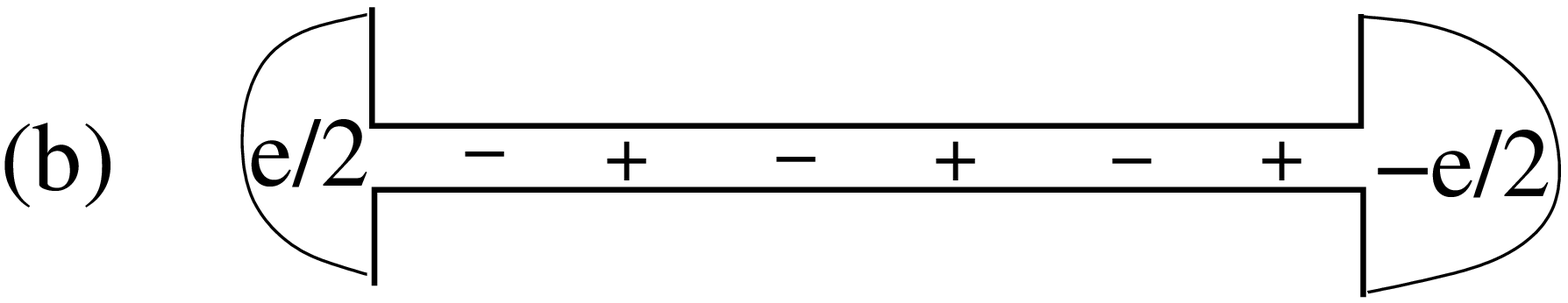} \hfill
\includegraphics[height=0.063\textheight]{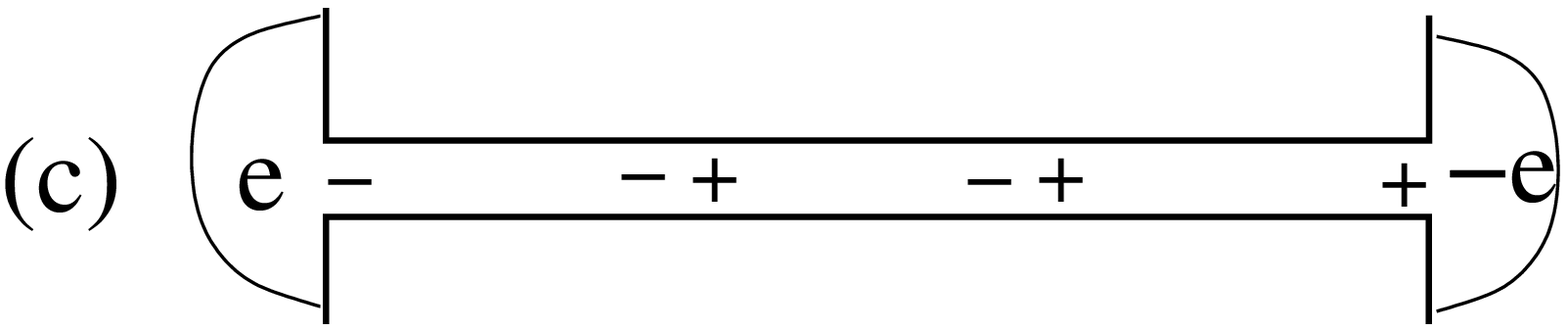}
\end{center}
\caption{ Three consecutive states of the channel with three
pairs. (a) Compact pairs with $ q = -q'\approx 0$, (b) the saddle
point state with $ q = -q'= 1/2$, (c) Compact pairs with $ q
=-q'\approx 1$. Other saddle points may be obtained e.g. by moving
the two same sign ions away from each other.} \label{fig5}
\end{figure}

One can repeat this calculation for an arbitrary number of pairs
inside the channel. One can see that the maximum energy state, the
system goes through to contribute to the charge transport, is
always reached when $q=\pm 1/2$ and always has energy $U_L(0)$. We
shall refer to such a state with the half-integer boundary charge
as the {\em collective saddle point} configuration.
Fig.~\ref{fig5} illustrates how the channel with three originally
compact pairs oriented along the external electric field transfers
a unit charge. It starts from the state with  $q = -q'\approx 0$,
goes to the top of the barrier with $q = -q'= 1/2$ and ends up
again in the state of compact pairs  with $ q =- q'\approx 1$. The
net result of such a process is a transfer of the unit charge
across the channel.

From the fact that the  energy needed to transfer the charge is
completely independent of the ion concentration one may tend to
conclude that the same is true regarding the activation barrier.
Such a conclusion is premature, however. Indeed, the activation
barrier is given by the maximum of the {\em free} energy and thus
includes the (negative) entropy contribution. As we explain below,
the collective saddle point configuration possesses  a huge
degeneracy and thus has a rather large entropy. This entropy
reduces the activation barrier down to Eq.~(\ref{weakdop}).

For a more quantitative consideration one needs to discuss the
exchange of ions between the bulk reservoirs and the channel.
Although the concentration of single ions in the channel is
strongly depleted, compact neutral pairs enter the channel freely.
Thus, typically there are many interacting pairs of ions inside
the channel. We have to discuss separately their equilibrium
(minimum energy) state and their collective saddle point (barrier)
state when they contribute to the transport.  It is convenient to
distinguish  the range of relatively low concentrations $c$, when
$\alpha \ll 1$, and  the range of high concentrations, $\alpha >
1$. The concept of the boundary charges, introduced above, is
helpful in both cases.

\subsection{Short channel at low concentration of salt}
\label{sec32}

In this section we consider the case of  low concentration of ions
in the bulk  salt solution, such  that  $\alpha \ll 1$. A typical
concentration of pairs inside the channel is even lower:
$n_p=2n\alpha\ll n$. To transfer the charge across the channel the
plasma must go trough the collective  saddle point state. Such
saddle point is characterized by the half--integer boundary
charge, say $q=1/2;\, q'=-1/2$. Half--integer  boundary charges
apply the external electric field $E_0$ across the channel. All
dipoles of compact pairs turn in the direction of the field, so
that positive and negative ions alternate. One can see now that
inside each pair the electric field changes its sign and turns to
$-E_0$ (recall the analogy with the uniformly charged planes,
producing field $E_0$ in both directions). As a result, at the
points where charges are located the electric field switches
between $E_0$ and $-E_0$. The electrostatic energy of the channel
is still equal to $U_{L}(0)$ and does {\em not} depend on the
positions  of the charges. This makes the charges {\em free} (as
long as they are ordered in the alternating sequence) from being
connected into the compact pairs. We shall call such peculiar
state -- the ordered free plasma (OFP).

Since in the OFP state the individual charges are free to enter
the channel, one expects that the concentration of free charges
(rather than the much lower concentration of the compact  pairs)
is the same in the channel and in the bulk. (Due to the
restriction of alternation the actual concentration in the channel
happens to be half of the one in the bulk.) One may  say that in
the OFP state ($q=1/2;\, q'=-1/2$) the gas of the free ions in the
bulk expands into the channel. This leads to the growth of the
system's entropy and thus reduction of the activation barrier.

In order to calculate the entropy gain $\Delta S$ of the OFP state
we assume that from the total number of ions, $2N$, some $2\Delta
N$ enter into the channel's volume $\Delta V = \pi a^{\,2} L$. The
rest $2N - 2\Delta N$ ions stay in the bulk volume $V$. In the
bulk $N-\Delta N$ positive ions and $N-\Delta N$ negative ones
produce additive contributions to the entropy. In the channel,
however, the gas is ordered so that ions of different signs
alternate and cannot exchange their positions. Therefore, the
entropy is the same as that of $2\Delta N$ identical ions. Thus,
the total entropy of the system is:
\begin{equation} \label{entropytotal}
{S\over k_B}=2(N-\Delta N) \ln \frac{eV/l_0^3}{N-\Delta N}+
2\Delta N \ln \frac{e \Delta V}{2\Delta Nl_0^3}.
\end{equation}
where $l_0^3$  is the normalization volume. The optimal  value of
$\Delta N$ maximizing $S$ is
\begin{equation}
\Delta N= \frac{1}{2}\, \frac{\Delta V}{V}\, N  = {nL\over 2}
\label{numberin}
\end{equation}
and thus the entropy gain is:
\begin{equation}
\Delta S=k_{B}nL\, . \label{deltaS}
\end{equation}
Recalling that the free energy at the collective saddle point is
$U_L(\alpha)=U_L(0)-T\Delta S$, we arrive at Eq.~(\ref{weakdop}).
In this derivation we have neglected entropy change of the
equilibrium state compared to $\Delta S$, because $n_{p}\ll n$.
This simplification is equivalent to omitting corrections which
are of the order of $\alpha^2$ in Eq.~(\ref{weakdop}). This
calculation is justified as long as $\Delta N\gg 1$.

\subsection{Short channel at high concentration of salt}
\label{sec33}

At high concentration of ions in the channel when $\alpha =nx_T >
1$ the thermal pairs overlap. Thus, one may expect that the ions
screen each other and become essentially free. If this is the
case, the concentration of ions inside the channel equals to that
outside, i.e. the 1d concentration  is equal to $n$. It is natural
to approach such a dense plasma with the mean-field Debye-H\"uckel
approximation. Recalling the analogy with the charged planes, one
needs to calculate the screening of a plane by other planes. This
is exactly what happens when 3d plasma screens a plane. As a
result, the electric field of a unit charge at a distance $x$ is
expected to be:
\begin{equation}
E(x)= E_0\, e^{-|x|/x_D}\, , \label{DH}
\end{equation}
where the Debye length, $x_D$, is given by the standard
expression:
\begin{equation}
x_D= \left({k_BT\over8\pi c e^{2}}\right)^{1/2} =
a\left({k_BT\over 8 ne^{2}}\right)^{1/2}. \label{Debyerad}
\end{equation}
In terms of $\alpha$ one has: $x_D \sim n^{-1}\alpha^{1/2}\sim x_T
\alpha^{-1/2}$, so that $n^{-1} < x_D < x_T$ for $\alpha \gg 1$.

The interaction energy of the system may be written as
\begin{equation}
W= \frac{\kappa_1 a^{\,2}L}{8}\!\int\limits_{-\infty}^{\infty}
\!\! dE\,E^{2}\,{\cal P}(E)\, , \label{energy}
\end{equation}
where ${\cal P}(E)$ is the distribution function of the random
electric field in the channel.  For $\alpha \gg 1$ it is Gaussian:
\begin{equation}
{\cal P}(E)= (2\pi \langle E^{2} \rangle)^{-1/2}
\exp\left[-E^{2}/2\langle E^{2} \rangle\right] \label{Gauss}
\end{equation}
with the mean square average value $\langle E^{2} \rangle = 4nx_D
E_0^{2} = W/(\pi a^{2}L)$. Indeed, fluctuations of the electric
field are created by the fluctuations of number of positive and
negative charges in two adjacent  segments of the channel of the
length $x_D$ each. A typical fluctuation has opposite charges of
the order of $Q\sim (nx_D)^{1/2}$, which interaction  energy is
$k_BT$.  Thus, it creates the electric field of the order of
$2QE_0=2(nx_D)^{1/2}E_0$.

As was discussed above, the activation barrier is associated with
the (free) energy difference between the state with half--integer
and integer boundary charges. According to the Debye-H\"uckel
theory any boundary charge is completely screened at the Debye
length, $x_D$, and thus is  a local perturbation. Then the
aforementioned energy difference is not an extensive quantity and
is proportional to $x_D$ rather than the channel length, $L$, as
was the case in the low density limit. Such a qualitative change
in the scaling of the activation barrier may occur only as a
result of a phase transition at some $\alpha_c \simeq 1$, where
the $L$--dependent part of the activation barrier vanishes.
It is well--known, however, that the
thermodynamics of 1d Coulomb gas, being well described by the
Debye-H\"uckel theory, does not exhibit any phase transitions
\cite{Lenard,Edwards,Schultz}. Thus, one either has to explain how
the Debye-H\"uckel thermodynamics may be consistent with the
$L$--dependent activation barrier, or find a phase transition in
the activation barrier which is not seen in the equilibrium
thermodynamics. We show below how the former of these two
alternatives works. There is {\em no} phase transition in the
activation barrier. Instead, the $L$-dependent activation energy
decreases exponentially with increasing $\alpha$ at $\alpha
> 1$.

The crucial point is that the  Debye--H\"uckel approximation
misses the discreteness of charge. Since the charges are discrete,
the electric field along the channel may change only in the steps
of $\pm 2E_0$, at the points where the charges are located.
Therefore at any point in the channel, having the boundary charges
$q$ and $q'=-q$, the electric field may be equal to $2E_0(q+M)$.
Here $M$ is an integer, given by the total charge to the left of
the point in question. In this manner the information about the
boundary charges ``propagates'' through the entire  channel (in an
apparent contradiction to the Debye--H\"uckel prediction,
Eq.~(\ref{DH})).

\begin{figure}[ht]
\begin{center}
\includegraphics[height=0.17\textheight]{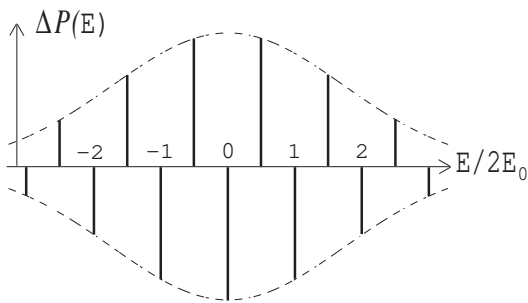}
\end{center}
\caption{ The function $\Delta {\cal P}(E)$ is the difference of
discrete distribution functions of electric field for $q=1/2$ and
$q=0$, which are made of set of $\delta$--functions weighted with
the envelope function ${\cal P}(E)$ given in Eq.~(\ref{Gauss}). }
\label{fignew}
\end{figure}

Let us show now that this ``quantization'' of the electric field
leads to the activation barrier proportional to $L$, though with
the exponentially small coefficient. To this end we employ the
distribution function of the electric fields, Eq.~(\ref{Gauss}),
as a weighting factor for the quantized values of the electric
field $2E_0(q+M)$. For example, at $q=0$ we sum over integers,
while at $q=1/2$ the sum is over half-integers. The transport
barrier $U_L(\alpha)= W(1/2)-W(0)$ is the difference between these
two sums, or in other words:
\begin{equation}
U_L(\alpha)=\frac{\kappa_1 a^{\,2}L}{8}
\int\limits_{-\infty}^{\infty}\!\! E^{2}\,\Delta {\cal P} (E) dE,
\label{newbarrie}
\end{equation}
where $\Delta {\cal P}$ is the difference of discrete distribution
functions for $q=1/2$ and $q=0$ illustrated in Fig.~\ref{fignew}.
It is clear that $U_L(\alpha)$ is exponentially small. Performing
summation over integer $M$ (instead of the integration, as in
Eq.~(\ref{energy})) with the help of the Poisson summation
formula, one finds the activation energy:
\begin{equation}
U_L(\alpha) = -\frac{\kappa_1 a^{\,2}L}{8}
\int\limits_{-\infty}^{\infty}\!\! dE\, E^{2}\,{\cal P}(E)\cos(\pi
E/E_0)\, . \label{cos}
\end{equation}
As a result, one finds for the activation barrier $U_L(\alpha)\sim
U_L\exp\{-\pi^2\sqrt{\alpha}\}$. Similarly to the dilute case the
activation barrier is proportional to the channel's length. Upon
increasing the salt concentration the barrier monotonously
decreases without undergoing the phase transition. Comparing the
exponent with the exact result, Eq.~(\ref{heavydop}), one notices
that the coefficient is slightly overestimated ($\pi^2$ instead of
$8$). The reason is that the fields $E$, contributing to the
integral in Eq.~(\ref{cos}), correspond to the charge of an
optimal fluctuation of the order $Q\approx enx_D$. Therefore both
assumptions of the applicability of the Gaussian distribution and
the theory of the Debye-H\"uckel linear screening to the optimal
charge fluctuations are only marginally correct. The quantitative
theory of the next section goes beyond these limitations. Unlike
the activation barrier, the thermodynamics is very weakly affected
by the ``quantization'' and is given by the Debye-H\"uckel theory
\cite{Lenard,Edwards} (up to exponentially small corrections).

\subsection{Long channel}
\label{sec34}

In the previous sections we dealt only with a short channel, where
the electric field does not escape from the channel to the media
with the small dielectric constant $\kappa_2$. For finite $\alpha$
the energy of the electric field is decreased due to the screening
by other ions. One may interpret this decrease of the activation
barrier as an enhancement of the dielectric constant: $\kappa_1\to
\kappa_1(\alpha)\equiv \kappa_1/f_0(\alpha)$. Indeed, from
Eq.~(\ref{short}), one may write
$U_L(\alpha)=e^{2}L/(2\kappa_1(\alpha)a^{\,
2})=f_0(\alpha)U_L(0)$.

Employing the concept of the concentration--dependent dielectric
constant, one may  formulate the results for the self-energy of an
ion in the long channel. Indeed, both the characteristic length of
the electric field decay, $\xi \to \xi(\alpha)$, and the
self--energy barrier, $U_\infty(0)\to U_\infty(\alpha)$, are
functions of the dielectric constant $\kappa_1$ (see
Eqs.~(\ref{criticaleq}), (\ref{long})), which depends now on the
concentration:  $\kappa_1\to \kappa_1(\alpha)$. The approximate
solution of Eq.~(\ref{criticaleq}) for $\xi(\alpha)$ is:
\begin{equation}
\xi(\alpha) \approx a\,\sqrt{\kappa_1(\alpha)\over \kappa_2}\,
\ln^{1/2}\left({\kappa_1(\alpha)\over \kappa_2}\right)\, ,
\label{criticalalpha}
\end{equation}
\begin{equation}
U_{\infty}(\alpha) = \frac{e^{2}}{a}\, \,
\frac{\ln^{1/2}(\kappa_1(\alpha)/\kappa_2)}
{\sqrt{\kappa_{1}(\alpha)\kappa_2}}. \label{longalpha}
\end{equation}
Since according to Eq.~(\ref{heavydop}), $f_0(\alpha)\sim
e^{-8\sqrt{\alpha}}$ for $\alpha> 1$, one finds
$\kappa_1(\alpha)\sim \kappa_1 e^{8\sqrt{\alpha}}$. According to
Eqs.~(\ref{criticalalpha}), (\ref{longalpha}) the decay length of
the field $\xi(\alpha)\sim e^{4\sqrt{\alpha}}$ grows exponentially
and the self--energy barrier is $U_\infty(\alpha)\sim
e^{-4\sqrt{\alpha}}$, cf Eq.~(\ref{heavydopalpha}).

For $\alpha \ll 1$ one must take into account deviations of $E(x)$
from $E_{0}$, cf. Eq.~(\ref{fielddecay}). As was discussed in
section \ref{sec32}, Eq.~(\ref{weakdop}) is based on the fact that
the field $E_{0}$ makes the pairs of positive and negative ions
free. A somewhat smaller field $E(x)$ leaves most of the pairs
bound. However, even bound pairs contribute to the entropy growth.
Let us consider a point charge in the middle of the channel with
the field decaying according to Eq.~(\ref{fielddecay}). It is easy
to see that the average arm of  pairs  grows for the dipoles
oriented along the direction of the electric field $E(x)$. For the
oppositely oriented pairs  the  arm shrinks. As a result, the
average pair length becomes $2 x_{T}(x) = 2 x_{T}
E_{0}^{2}/[E_{0}^{2}-E^{2}(x)]$ and for the local concentration of
pairs instead of Eq.~(\ref{npairs}) we find: $ n_{p}(x)= 2n^{2}
x_{T}(x)= n_{p} E_{0}^{2}/[E_{0}^{2}-E^{2}(x)]$. This  estimate is
valid for $|x|> \alpha\xi$, while at  smaller distances
$n_p\approx n/2$.
Thus the total number of pairs, $N_{p}$, brought into the channel
by the central charge may be estimated as:
\begin{equation}
\Delta N = 2 \int\limits_{\alpha \xi}^{L} [n_{p}(x)-n_{p}] dx =2
n\xi \alpha \ln(1/\alpha), \label{Npairs}
\end{equation}
The new pairs  add the entropic term $-k_B T \Delta N$ to the free
energy of the channel with the fixed central charge. This entropic
contribution leads to the suppression of the transport barrier
given by Eq.~(\ref{weakdopalpha}).

\subsection{Non-ohmic transport of the channel at large voltage}
\label{sec35}

Until now we dealt with the ohmic transport limited by small
voltages $eV\ll k_BT$. Let us consider the effect of a finite
voltage on the differential resistance in the opposite case, $eV >
k_BT$. We start from a short channel. As was discussed above, the
top of the transport barrier corresponds to the boundary charges
$\pm 1/2$. These charges are provided by the potential source
which makes the work $eV/2$. Thus, the conductance barrier becomes
lower by this energy and we arrive at Eq.~(\ref{diffconductance}).

In the long channel the electric field lines of a moving charge
leave the channel, go to the low dielectric constant media  and
eventually return to the water solution far from the channel.
Thus, each moving charge creates the two opposite sign images in
the two water--filled  semi--spaces. These images are similar to
the boundary charges. Attraction to each of them follows the three
dimensional Coulomb law. For example, a charge with the coordinate
$x$ close to $-L/2$ is attracted to the image with the energy
$-e^{2}/[4\kappa_{2}(L/2+x)]$, if $L/2 + x \gg \xi$. The sum of
the attraction energies to both water semi-spaces has maximum at
$x=0$. Thus, the top of the barrier corresponds to the charge
located at the center. In this position the image charges are
equal $q=q'=1/2$. If the applied electric field $V/L$ is smaller
than the field of each image in the center, $2e/(\kappa_{2}L^2)$,
the field only weakly affects position of the saddle point. At
such conditions the voltage only reduces the barrier by $eV/2$.
Thus, for a long channel one gets similarly to
Eq.~(\ref{diffconductance}):
\begin{equation}
R(V)\sim \exp\left\{\frac{U_{\infty}(\alpha)-eV/2}{k_BT}\right\}\,
\label{diffconductancelong}
\end{equation}
On the other hand, if $V/L \gg 2e/(\kappa_{2}L^2)$ the shape of
the barrier changes drastically. It looses the symmetry around
$x=0$. The saddle point moves to the distance $l \ll L/2$ from the
left border. The length $l$ is determined by the maximum of the
sum of the attractive Coulomb potential of the charge to the left
image and the linear potential of the uniform electric field
$V/L$. This maximum is located at the distance $l =
(eL/4\kappa_{2}V)^{1/2}$. As a result, the barrier  is reduced  by
$2eVl/L = (e^{3}V/\kappa_{2}L)^{1/2}$, thus we arrive at
Eq.~(\ref{Frenkel}). This result was first derived by
Frenkel~\cite{Frenkel} for the field--induced ionization of
Coulomb impurities in semiconductors.

The longer the channel, the smaller is the range of applicability
of Eq.~(\ref{diffconductancelong}) and the larger is  the range of
applicability of Eq.~(\ref{Frenkel}). Notice that the applied
voltage $V$ does not interfere with the effect of the salt on the
barrier height. In the Eq.~(\ref{diffconductance}),
Eq.~(\ref{diffconductancelong}) and Eq.~(\ref{Frenkel}) the
concentration--dependent height of the barrier determines only the
voltage at which the activation exponent is of order one.
Equations~(\ref{diffconductance}), (\ref{diffconductancelong}) and
(\ref{Frenkel}) are valid only at smaller voltages.

\section{Thermodynamics of the short channel}
\label{sec4}

\subsection{Partition function}
\label{sec41}

In this section we focus on a relatively short channel, such that
the electric field lines do not escape to the media with the small
dielectric constant. The calculation of the partition function is
similar to that in Ref.~\cite{Edwards}. We briefly reproduce it
here to focus on the role of the boundary charges and to
generalize it later to the short channel case.  The interaction
energy of plasma consisting of $N$ positive and $N'$ negative
charges is given by
\begin{equation}
  U={1\over 2} \sum\limits_{i,j=1}^{N+N'} \sigma_i \sigma_j \Phi(x_i-x_j)\, ,
                                    \label{totalenergy}
\end{equation}
where $\sigma_j=1$ for $j=1,\ldots N$ and $\sigma_j=-1$ for
$j=N+1,\ldots N+N'$. The interactions are mediated through the 1d
Coulomb potential:
\begin{equation}
\Phi(x)=\Phi(0) - eE_0|x|
                                 \label{potential}
\end{equation}
Here $\Phi(0)$ is an artificial constant (twice of the ion's
self--energy) which is  used  to impose charge neutrality by
taking $\Phi(0)\to \infty$ limit. To this end  the diagonal terms,
representing the self-energy of the charges are included  in
Eq.~(\ref{totalenergy}).

Following  discussion of Section \ref{sec31}, one also introduces
two boundary (in general non-integer) charges $q$ and $q'$ at the
end points of the channel $x=\mp L/2$. With these two charges
included the total energy takes the form:
\begin{equation}\label{totalenergy_q}
U(q,q')={1\over 2}\int\!\!\!\!\!\!\int\limits_{-L/2}^{L/2}\!\!
dxdx' \rho(x)\Phi(x-x')\rho(x')\, ,
\end{equation}
where the charge density is defined as
\begin{equation}\label{chargedensity}
  \rho(x)\equiv \!\!\sum\limits_{j=1}^{N+N'}\!\! \sigma_j
  \delta(x-x_j)+q\delta(x+L/2)+q'\delta(x-L/2)\, .
\end{equation}

We are interested in the grand-canonical partition function of the
gas defined as
\begin{eqnarray}\label{partition}
  && Z_L(q,q') = \sum\limits_{N,N'=0}^\infty e^{\mu(N+N')/(k_BT)}{1\over
  N!N'!} \\
&&\times \prod\limits_{j=1}^{N+N'}\left({\pi a^{\,2}\over
l_0^2}\int\limits_{-L/2}^{L/2} {dx_j\over l_0} \right) e^{-{1\over
k_B T} U(q,q')} \, ,\nonumber
\end{eqnarray}
where $\mu$ is the chemical potential of the charges (the same for
pluses and minuses) and $l_0$ is a microscopic scale related to
the bulk salt concentration as $c=e^{\mu/k_BT}/l_0^{3}$. Factor
$\pi a^2/l_0^2$ originates from the integrations over transverse
coordinates. Since we have in mind the viscous dynamics of the
ions, rather than the inertial one, the kinetic energy is not
included in the partition function.

Formally any configuration of $N$ pluses and $N'$ minuses is
allowed in this expression. One expects, though, that only neutral
configurations with $N-N'+q+q'=0$ provide non--zero contributions
due to the presence of the large self--energy, $\Phi(0)$. As a
result, one expects that $Z_L(q,q')\sim \delta_{q',m-q}$, where
$m=N'-N$ is an integer number. We shall see later that this is
indeed the case.

To proceed with the evaluation of $Z_L(q,q')$ we introduce the
resolution of unity written in the following way:
\begin{widetext}
\begin{eqnarray}\label{res_unity}
  1&=&\int\!\!{\cal D}\! \rho(x)\,\, \delta\!\!\left(\rho(x) -
\sum\limits_{j=1}^{N+N'} \sigma_j
  \delta(x-x_j)-q\delta(x+L/2)-q'\delta(x-L/2)\right)
  \nonumber \\ &=&
\int\!\!{\cal D}\! \rho(x)\int\!\!{\cal D} \theta(x)\,\,
   \exp\left\{\,-i\!\!\!\int\limits_{-L/2}^{L/2} \!dx\, \theta(x)\left(\rho(x) -
\sum\limits_{j=1}^{N+N'} \sigma_j
  \delta(x-x_j)-q\delta(x+L/2)-q'\delta(x-L/2)\right)\right\}\\
  &=&\int\!\!\!\int{\cal D}\! \rho(x){\cal D} \theta(x)\,\,
   \exp\left\{\,-i\left(\int\limits_{-L/2}^{L/2}  \!dx\, \theta(x)\rho(x)  -
\sum\limits_{j=1}^{N+N'} \sigma_j
  \theta(x_j)-q\theta(-L/2)-q'\theta(L/2)\right)\right\}~.\nonumber
\end{eqnarray}
Substituting this identity into the expression for the partition
function, one finds:
\begin{eqnarray}\label{partition1}
  Z(q,q')=&&\int\!\!\!\! \int\limits_{-\infty}^{\infty} \frac{d\theta_i
  d\theta_f}{4\pi^2}\,\, e^{ iq\theta_i+iq'\theta_f }
  \int\!\!\!\int{\cal D}\! \rho(x){\cal D} \theta(x)\,
   \exp\left\{-{1\over 2k_BT}\int\!\!\!\int\limits_{-L/2}^{L/2} \!\! dxdx'
\rho(x)\Phi(x-x')\rho(x')
  - i\int\limits_0^L \!dx\, \theta(x)\rho(x)\right\}\nonumber \\
   && \times\sum\limits_{N=0}^\infty {1\over N!}\, e^{\mu N/(k_BT)}
\left({\pi
a^{\,2}\over l_0^3} \int\!\!
   dx \, e^{i\theta(x)} \right)^N
   \times \sum\limits_{N'=0}^\infty {1\over N'!}\, e^{\mu N'/(k_BT)} \left({\pi
a^{\,2}\over l_0^3}\int\!\!
   dx \, e^{\,-i\theta(x)} \right)^{N'}\\
 = && \int\!\!\!\! \int\limits_{-\infty}^{\infty} \frac{d\theta_i
  d\theta_f}{4\pi^2}\,\, e^{ iq\theta_i+iq'\theta_f }\int\!\! {\cal D}
\theta(x)\,\,
   \exp\left\{-{k_BT\over 2} \int\!\!\!\int\limits_{-L/2}^{L/2} \!\! dxdx'
\theta(x)\Phi^{-1} (x-x')\theta(x')
  + 2\pi a^{\,2}c \int\limits_{-L/2}^{L/2}  \!dx
  \cos \theta(x)\right\} \, .\nonumber
\end{eqnarray}
\end{widetext}
 The integral over $\theta(x)$ runs over all
functions with the boundary conditions $\theta(-L/2)=\theta_i$ and
$\theta(L/2)=\theta_f$. It is straightforward to verify that
$\Phi^{-1}$ operator is  given by $\Phi^{-1}(x-x')=-(2eE_0)^{-1}
\delta(x-x') \partial^2_x$. As a result, the functional integral
on the r.h.s. of the last expression takes the form
\begin{equation}\label{greenfunct}
  G_L(\theta_i,\theta_f)\equiv
  \int\!\! {\cal D}  \theta(x)\,\,
  e^{- {x_T\over 2}\!  \int\!\! dx\left[{1\over 2} (\partial_x \theta)^2 -
{4\alpha\over
 x_T^2} \cos \theta(x)\right] }\, ,
\end{equation}
where $x_T$ and $\alpha$ are given by Eqs.~(\ref{thermallength})
and (\ref{alpha}).

Expression (\ref{greenfunct}) represents the ``quantum
mechanical'' probability to propagate from the point $\theta_i$ to
the point $\theta_f$ during the (imaginary) ``time'' $L$. The
corresponding  ``Schr\"odinger equation''  for the
``wave-function'' $\Psi=\Psi(\theta,x)$ has the form:
\begin{equation}\label{shrodinger0}
  -{1\over 2}
 \frac{\partial^2 \Psi}{\partial \theta^2} - \alpha
\cos
 \theta\,\, \Psi = - {x_T\over 2}{ \partial\Psi\over\partial x} \, .
\end{equation}
One may look for the eigenfunctions of this equation:
$\Psi_k(\theta,x)=\Psi_k(\theta)\exp\{-2\,\epsilon_k x/x_T\}\,$,
labelled by a quantum number $k$. The corresponding
``Schr\"odinger equation''  for the stationary eigenfunctions has
the form:
\begin{equation}\label{shrodinger}
  -{1\over 2}
 \frac{\partial^2 \Psi_k(\theta)}{\partial \theta^2} - \alpha
\cos
 \theta\,\, \Psi_k(\theta) = \epsilon_k\Psi_k(\theta)\, ,
\end{equation}
where $\epsilon_k$ is the  energy. In terms of the stationary
eigenfunctions  the propagator takes the  form:
\begin{equation}\label{greenfunct1}
  G_L(\theta_i,\theta_f) = \sum\limits_k
  \Psi_k(\theta_i)\bar\Psi_k(\theta_f)\, e^{-2\epsilon_k L/x_T}\, .
\end{equation}
Finally, the partition function, Eq.~(\ref{partition1}), is
nothing but the propagator in the momentum representation and thus
may be written as:
\begin{equation}\label{partition2}
  Z_L(q,q') = G_L(q,-q')=\sum\limits_k
  \Psi_k(q)\bar\Psi_k(-q')\, e^{-2\epsilon_k L/x_T}\, ,
\end{equation}
where $\Psi_k(q)\equiv \int d\theta/(2\pi) \Psi_k(\theta)
\exp\{i\theta q\}$. Relation between the thermodynamics of the 1d
Coulomb gas and the Mathieu  equation (\ref{shrodinger}) was
realized long ago by Lenard~\cite{Lenard}. He was interested,
however, only in the equilibrium thermodynamics, rather than the
transport characteristics.

\subsection{Green function}
\label{sec42}

One can employ now the well-known results for the Schr\"odinger
equation in the periodic potential, Eq.~(\ref{shrodinger}), to
find the propagator, $G_x(q,\tilde q)$. The wave-functions are
given by the Bloch-waves:
\begin{equation}\label{Bloch}
  \Psi_k^{(j)}(\theta) = e^{-ik\theta}u_k^{(j)}(\theta) =
 e^{-ik\theta}\sum\limits_{m=-\infty}^\infty
 u_{k,m}^{(j)}e^{-im\theta} \,,
\end{equation}
where $|k|< 1/2$ is the quasimomentum and $u_k^{(j)}(\theta)$ is
the period-$2\pi$ (Bloch) function in the $j$-th band. In the last
expression one has expanded the Bloch function in the Fourier
series. The Fourier transform of the eigenfunction with the
quantum numbers $k,(j)$ is therefore given by
\begin{equation}\label{Fourier}
  \Psi_k^{(j)}(q) =
 u_{k,m}^{(j)} \delta(q-k-m) \, ,
\end{equation}
where $m=q-k$ must be an integer, while $|k|<1/2$; i.e. $k=[q]$,
where the square brackets denote the fractional part. We found,
thus, that the propagator conserves momentum up to an integer $m$
(umklapp processes): $q = \tilde q+m$. This is a manifestation of
the charge neutrality. The integer parts of $q$ and $q'=-\tilde q$
may be arbitrary. Since the integer parts may be easily screened
by attracting the corresponding number of counter-ions from the
solution, they do not play an important role in the subsequent
discussion. One finds then for the Green function:
\begin{equation}\label{partition3}
  G_x(q,\tilde q) = \delta([q]-[\tilde q])
 \sum\limits_{j} u_{[q],m}^{(j)} \bar u_{[q],\tilde m}^{(j)}\,
e^{-2\epsilon_q^{(j)}
 x/x_T}\, ,
\end{equation}
where we have used the fact that the band dispersion
$\epsilon_k^{(j)}$ is a periodic function of the quasimomentum.
The common fractional part of the boundary charges plays the role
of the Bloch quasimomentum. It is important to notice that if $q$
is fixed, only the single state from each Bloch band contributes
to the partition function.

For a sufficiently long channel, only the lowest Bloch band,
$j=0$, should be retained in this expression, while the upper
bands provide  exponentially small corrections. Since the free
energy is given by the logarithm of the propagator, one may also
disregard  $u_{q,m}^{(0)}$ factors. Indeed, the latter contribute
to the free energy only a $L$--independent (boundary) term, which
is relatively small if $L\gg x_T$. Therefore, the propagator may
be approximated by:
\begin{equation}\label{green3}
  G_x(q,\tilde q) \approx  \delta([q]-[\tilde q])\,
e^{-2\epsilon_q^{(0)}x/x_T}\, ,
\end{equation}
where the $\epsilon_q^{(0)}(\alpha)$ is the dispersion relation of
the lowest Bloch--Mathieu band. Finally, the extensive part of the
free energy of the channel according to Eq.~(\ref{partition2}) is
given by:
\begin{equation}\label{freeenergy}
  F_L(q)\equiv -k_BT\ln Z_L(q,q')= k_BT\,\frac{2L }{x_T}\,\epsilon^{(0)}_q\, ,
\end{equation}
provided $[q']=-[q]$. The fact that the free energy is a periodic
function of $q$ with the unit period reflects the perfect
screening of the integer part of the boundary charge.

\subsection{Lowest Bloch band}
\label{sec43}

Since the free energy is given by the dispersion relation
$\epsilon_q^{(0)}(\alpha)$, let us discuss this function in more
details. We concentrate on the first Brillouin zone: $|q|<1/2$.

In the extreme low concentration limit, $\alpha\to 0$, the
dispersion relation is given by $\epsilon^{(0)}_q=q^2/2$. In the
dilute limit, $\alpha \ll 1$, the cosine potential in
Eq.~(\ref{shrodinger}) may be treated in the perturbation theory.
Save for the points $q=\pm 1/2$, there are no first order
corrections. The second order perturbation theory leads to:
\begin{equation}
\epsilon^{(0)}_q\approx {q^2\over 2} -\frac{\alpha^2}{1-4q^2}\, .
                            \label{seconorder}
\end{equation}
The divergency at $q=\pm 1/2$ signifies that the narrow interval
around  these two points, $|q\mp 1/2|\lesssim \alpha \ll 1$,
requires a special treatment. Indeed, in the narrow region around
$q= \pm 1/2$ points the energy gap opens up. Linearizing the bare
spectrum $\,q^2/2\,$ in the vicinity of these points, one reduces
the Schr\"odinger equation to the spectral problem for the
$2\times 2$ matrix Hamiltonian:
\begin{equation}\label{2times2}
  {1\over 2} \left(\begin{array}{cc}
  {1\over 4} +{\delta q} & {\alpha}\\
  {\alpha} & {1\over 4} -{\delta q}
  \end{array}\right)\, ,
\end{equation}
where $\delta q\equiv 1/2 \pm q$. Finding the spectrum, one
obtains that the lowest Bloch band at $|\delta q| \lesssim\alpha$
takes the form:
\begin{equation}\label{point1/2}
  \epsilon^{(0)}_q\approx {1\over 8} \left(1-4\sqrt{\alpha^2 +\delta q^2}\right)
\,.
\end{equation}
At $q=\pm 1/2$ the energy reaches its maximum
$\epsilon^{(0)}_{1/2}=(1-4\alpha)/8$.

In the high concentration limit, $\alpha> 1$,  the amplitude of
the cosine potential is large.  The band is narrow and is centered
near the ground-state  of an isolated (almost) parabolic potential
well:
\begin{equation}\label{highT}
\epsilon_q^{(0)} \approx - \alpha + {1\over 2} \sqrt{\alpha} \, .
\end{equation}
The momentum dispersion originates from the tunnelling between the
adjacent wells. One may treat it in the tight--binding
approximation and find for the dispersion relation:
\begin{equation}\label{dispersion}
\epsilon_q^{(0)} = -\alpha +{\sqrt{\alpha}\over 2} -
2^{-4}f_0(\alpha)\cos (2\pi q)\, .
\end{equation}
The band-width, $2^{-3}f_0(\alpha)$, may be calculated in the WKB
approximation (for the numerical factor see Ref.~\onlinecite{AS}):
\begin{equation}\label{bandwidth}
  2^{-3} f_0(\alpha)= 2^{4} \pi^{-1/2} \alpha^{3/4} \exp\left\{-
8\sqrt{\alpha}\,\right\}\,.
\end{equation}

\subsection{Thermodynamics}
\label{sec44}

One can  employ now Eq.~(\ref{freeenergy}) to extract the
thermodynamic characteristics of the 1d Coulomb plasma (see also
Ref.~\onlinecite{Lenard}). In particular, we are interested in the
salt concentration inside the channel and the average interaction
energy. The former is given by:
\begin{equation}\label{concentration}
n_p=-{1\over 2 L}{\partial F\over \partial \mu} = -{1\over 2
L}{\partial F\over \partial \alpha}{\partial \alpha\over \partial
\mu} = -n\,{\partial \epsilon_q^{(0)}(\alpha)\over
\partial \alpha}\, ,
\end{equation}
while the interaction energy is:
\begin{equation}\label{interactionenergy}
  W=-T^2{\partial\over \partial T}\left({F\over
  T}\right)=-2nLk_BT\alpha {\partial\over \partial \alpha}\left(
  {\epsilon_q^{(0)}(\alpha)\over\alpha}\right) \, .
\end{equation}

For  low  concentration, $\alpha\ll 1$, and $|q\mp 1/2|> \alpha$
the pair concentration is: $n_{p}(q)=2n\alpha/(1-4q^2)$. This
means that as long as $|q\mp 1/2| > \alpha$, the  pair
concentration inside  the channel is low $n_{p}\ll n$. In terms of
$n_p$ the free energy takes the simple form: $F(n_p) =
\mbox{const}-Ln_p k_BT$, corresponding to the ideal gas of dipole
pairs. The interaction energy is given by $W=eE_0Lq^2+ n_pLk_BT$,
showing that there is energy $k_BT$ per pair (in addition to the
interaction energy of the boundary charges).

Consider now the narrow region around $q= \pm 1/2$ points, where
$|q\pm 1/2|\lesssim \alpha$. Employing Eq.~(\ref{point1/2}), one
finds for the pair concentration: $n_p(1/2) =n/2\gg n_p(0)$. In
fact, it is only half of the naive non--interacting estimate for
the pair concentration inside the channel: $n$. This is the OFP
state, where the ions are free to move as long as they are ordered
in the alternating sequence. Indeed, for the free energy one finds
$F(n_{p}) = \mbox{const}-2Ln_p k_BT$, indicating an ideal gas of
individual ions. For $q=1/2$ one has $W=eE_0L(1/2)^2$, meaning
that the entire interaction energy comes from the boundary charges
and not from the inside pairs. Notice that slightly away from the
collective saddle point, $q=1/2$, the plasma in the channel is
strongly non--ideal.

In the dense limit $\alpha >1$ the thermodynamic properties are
almost independent from the boundary charges, $q$. Employing
Eq.~(\ref{highT}), one finds: $n_p= n-n/(4\sqrt{\alpha})$. The
concentration in the channel is almost equal to the bulk
concentration. For the interaction energy one finds
$W=nLk_BT/(2\sqrt{\alpha})$ in exact agreement with the
Debye-H\"uckel calculation, cf. Eq~(\ref{energy}). Since the
interaction energy per ion is much less than the temperature, the
thermodynamics of the plasma is almost identical to that of the
ideal gas of free ions \cite{Lenard}. Despite of this, the plasma
possesses a thermodynamically extensive ($\propto L$) activation
barrier for the charge transport.

\subsection{Activation barrier}
\label{sec45}

To calculate the activation barrier, one needs to find the free
energy of the collective saddle point state. To this end we
introduce an additional  unit charge (say negative) at some point,
$x_0$, inside the channel. Due to the perfect charge neutrality
the system must respond by bringing two screening boundary
charges: $q$ and $q'=1-q$, where $q$ depends on $x_0$. An example
of such dependence is discussed above, see Eq.~(\ref{parabolic}).
One can invert the logic and claim that by changing the boundary
charge $q$ one changes the equilibrium position of the test
charge, $x_0$. It is clear that by varying $q$ from zero to one,
the negative charge is moved from $x_0=L/2$ to $x_0=-L/2$. The
activation barrier for such process is determined by the maximum
of the free energy as  function of $q$. According to
Eqs.~(\ref{partition2}), (\ref{green3}) and (\ref{freeenergy}) the
partition function  is given by:
\begin{equation}\label{Zq1-q}
  Z_L(q,1-q)=G_L(q,q-1)=e^{-2\epsilon^{(0)}_qL/x_T}\, .
\end{equation}
Thus the free energy  reaches the maximum at $q=1/2$, which
corresponds to the test charge being in the middle of the channel.
The minima are reached at $q=0$ and $q=1$, corresponding to the
test charge being at $x_0=\pm L/2$.

Finally, the activation barrier is given by:
\begin{equation}\label{actbarrier1}
  U_L(\alpha)= F_L(1/2)-F_L(0) =
  U_L(0)f_0(\alpha)\, ,
\end{equation}
where the bare barrier $U_L(0)$ is given by Eq.~(\ref{short}) and
the universal function $f_0(\alpha)$ is proportional to the Bloch
band-width:
\begin{equation}\label{falpha}
  f_0(\alpha) =
  8\left(\epsilon^{(0)}_{1/2}(\alpha)-\epsilon^{(0)}_0(\alpha)\right)\,.
\end{equation}
This function is plotted in Fig.~\ref{fig3}. The two limiting
cases of low and high concentration, Eqs.~(\ref{weakdop}) and
(\ref{heavydop}), follow immediately from Eqs.~(\ref{point1/2})
and (\ref{bandwidth}), correspondingly.

\section{Long  and intermediate channels}
\label{sec5}

\subsection{Partition function}
\label{sec51}

In the long channel $L> \xi> x_T$, where the characteristic length
$\xi$ is given by Eq.~(\ref{criticaleq}), the electric field lines
escape to the media with the small dielectric constant,
$\kappa_2$. As a result, the interaction potential gradually
looses it 1d character and crosses over to the 3d Coulomb law
$\Phi(x)\sim 1/(\kappa_2x)$. According to Eq.~(\ref{fielddecay})
in the large range of distances it has the form:
\begin{equation}
\Phi(x)=eE_0\xi\,e^{-|x|/ \xi}\, .
                                            \label{modelpotential}
\end{equation}
Our previous consideration, c.f. Eq.~(\ref{potential}),
corresponds to the limit $\xi\to\infty$. Since
$\Phi(0)=eE_0\xi\equiv 2U_\infty(0)$ (c.f. Eq.~(\ref{long})) is
now finite, there is no reason to expect the strict charge
neutrality. It is straightforward to invert
Eq.~(\ref{modelpotential}): $\Phi^{-1}(x-x')=(2eE_0)^{-1}
\delta(x-x') (\xi^{-2}-\partial^2_x)$. Thus, one can repeat the
mapping onto the quantum mechanics and arrive at the following
Schr\"odinger equation for the wave-function $\Psi(\theta,x)$:
\begin{equation}\label{shrodinger2}
  -{1\over 2}
 \frac{\partial^2 \Psi}{\partial \theta^2} - \alpha \cos
 \theta\,\, \Psi +  {1\over 2} \left({x_T\over 2\xi}\right)^2 \theta^2\, \Psi=
 -{x_T\over 2}{\partial \Psi\over \partial x}\, .
\end{equation}
As before, the grand--canonical partition function, $Z_L(q,q')$,
of the plasma with the fixed boundary charges is  given by
Eq.~(\ref{partition2}):
\begin{equation}\label{partition4}
  Z_L(q,q') = G_L(q,-q')=\sum\limits_k
  \Psi_k(q)\bar\Psi_k(-q')\, e^{-2\epsilon_k L/x_T}\, .
\end{equation}
This is nothing but the Green function of Eq.~(\ref{shrodinger2})
in the momentum representation.

We shall show now that one may use the small parameter
$x_T/(2\xi)\ll 1$ to find the Green function $G_x(q,\tilde q)$ in
the quasiclassical approximation. The results obtained this way
are restricted  to the concentration range $\alpha>
\sqrt{x_T/\xi}$. For lower concentrations one may use the standard
first order perturbation theory in $\alpha$ for
Eq.~(\ref{shrodinger2}). It leads to the linear in $\alpha$
correction to the activation barrier, which is bound to be smaller
than $k_BT$ if $\alpha < \sqrt{x_T/\xi}$. By this reason we shall
not discuss it here.

To proceed we rewrite the Schr\"odinger equation
(\ref{shrodinger2}) in the basis of the Bloch waves,
$\Psi_k^{(j)}(\theta)$, Eq.~(\ref{Bloch}). This basis diagonalizes
the first two terms on the l.h.s. of Eq.~(\ref{shrodinger2}). The
only non--diagonal term in this basis is the confinement potential
proportional to $\theta^2$. In the Bloch basis it takes the form
$-\delta_{k,k'}[\delta^{j,j'}\partial^2_k
+2v^{j,j'}_1(k)\partial_k +v^{j,j'}_2(k)]$, where
$v_{s}^{j,j'}(k)=\sum_m \bar u_{k,m}^{(j)}\partial_k^s
u_{k,m}^{(j')}$ and $s=1,2$. Fortunately, one does not need to
evaluate functions $v_s^{j,j'}(k)$. Indeed, due to the small
coefficient $x_T^2/(8\xi^2)\ll 1$ in front of the $\theta^2$ term,
only the highest  derivative, $\sim \partial^2_k$ contributes to
the leading order in $\xi/x_T$ of the WKB action. By the same
reason one may work only with the lowest band, $j=0$, and
disregard differences between $k$ (Bloch quasimomentum) and $q$
(momentum). As a result, Eq.~(\ref{shrodinger2}) takes the form:
\begin{equation}\label{shrodinger3}
  -{1\over 2} \left({x_T\over 2\xi}\right)^2
 \frac{\partial^2 \Psi}{\partial q^2} +\epsilon_q^{(0)} \Psi=
 -{x_T\over 2}{\partial \Psi\over \partial x}\, .
\end{equation}
Here $\Psi=\Psi(q,x)$ satisfies  the periodic boundary conditions
imposed at $q=\pm 1/2$. This is the Schr\"odinger equation for a
particle with the heavy mass $(2\xi/x_T)^2\gg 1$ in the potential
$\epsilon_q^{(0)}(\alpha)$.  The latter is the dispersion relation
of the lowest Bloch band of the Mathieu equation
(\ref{shrodinger}). Thanks to the large mass one can employ the
WKB approximation  to evaluate the Green function,
Eq.~(\ref{partition4}). In this approximation the Green function
is given by the exponentiated {\em classical} action corresponding
to Eq.~(\ref{shrodinger3}). Such an action may be written in terms
of the Lagrangian classical ``coordinate'' $q(x)$ as:
\begin{equation}\label{classaction0}
  S_L(q,\tilde q)={2\over x_T} \int\limits_{-L/2}^{L/2}\!\!\! dx \left[
{\xi^2\over 2}(\nabla_x q(x))^2 +
  \epsilon_{q(x)}^{(0)}\right]\, ,
\end{equation}
where $q(-L/2)=q$ and $q(L/2)=\tilde q$. A useful analogy is that
of a 1d string with the displacement $q(x)$ and rigidity $\xi^2$,
placed in the periodic potential profile $\epsilon^{(0)}_q$. The
optimal configuration of such string is given by the minimum of
the Lagrangian action (\ref{classaction0}):
\begin{equation}\label{classequation}
  \xi^2\nabla^2_x q - {\partial \epsilon_q^{(0)}\over \partial q}
  =0\, ,
\end{equation}
which must be solved with the specified boundary conditions. The
action is then given by:
\begin{equation}\label{classaction1}
  S_L(q,\tilde q)={2\over x_T}\left[L {\cal E} +
  \xi\int\limits_q^{\tilde q}\!\! p(q) \, dq\right]\, ,
\end{equation}
where the canonical classical momentum is given by
$p(q)=\sqrt{2(\epsilon_q^{(0)}-{\cal E})}$ and ${\cal E}$ is the
integral of motion, found from Eq.~(\ref{classequation}) and the
boundary conditions. Finally, the free energy of the 1d plasma
with the boundary charges $q$ and $q'$ is given by:
\begin{equation}\label{classaction}
  F_L(q,q')=-k_BT\ln G_L(q,-q') = k_BT S_L(q,-q')\, .
\end{equation}

In the equilibrium state $q=q'=0$. The string lies in the minimum
of the potential: $q(x)=0$ and thus ${\cal E}=\epsilon_0^{(0)}$,
while $p(q)=0$.  The corresponding action is
$S_L(0,0)=2L\epsilon_0^{(0)}/x_T$ and  the equilibrium free energy
is given by:
\begin{equation}\label{freeshort}
  F_L(0)= k_BT\, {2L\over x_T}\, \epsilon_0^{(0)}\,.
\end{equation}
This is the same result we found for the short channel,
Eq.~(\ref{freeenergy}). Thus the equilibrium thermodynamics is not
affected by the escape of the electric field from the channel (for
not too low concentration: $\alpha> \sqrt{x_T/\xi}$). On the
contrary, the activation barrier is essentially different.

\subsection{Activation barrier}
\label{sec52}

To calculate the activation barrier consider again a negative unit
test charge at some point $x_0$ inside the channel. It induces the
screening charges $q$ and $q'$ at the boundaries. Since the saddle
point is reached when the test charge is placed in the middle of
the channel, we may put $q'=1-q$. Unlike the short channel case,
there is no strict charge neutrality. Therefore we should not
expect $q$ to be $1/2$ at the saddle point. Instead one should
consider a function $q(x)$ with the boundary conditions
$q(-L/2)=q$ and $q(L/2)=1-q$, which is a solution of
Eq.~(\ref{classequation}). From the symmetry one expects
$q(0)=1/2$, corresponding to the test charge being at $x_0=0$. For
the free energy of the collective saddle point one thus obtains:
\begin{equation}\label{freecsp}
  F_L^{sp}(q)=k_BT\, S_{L}(q,1-q)\,,
\end{equation}
where $S_L(q,1-q)$ is the action (\ref{classaction0}) on the
optimal string configuration and the initial displacement, $q$, is
determined from the condition that the  string ``slides'' from
$q(0)=1/2$ down to $q(-L/2)=q$ according to
Eq.~(\ref{classequation}).

In the short channel limit, $L\ll \xi$,  the string does not bend
and thus $q(x) = 1/2$ at any point. As a result ${\cal
E}=\epsilon_{1/2}^{(0)}$ and the classical action is given by the
first term on the r.h.s. of Eq.~(\ref{classaction1}):
$S_{L}(q,1-q)\to S_{L}(1/2,1/2)= 2L\epsilon_{1/2}^{(0)}/x_T$. The
activation barrier is thus given by Eqs.~(\ref{actbarrier1}) and
(\ref{falpha}), as expected.

In the long channel limit, $L\gg \xi$, the string goes from the
minimum at $q=0$ to the neighboring minimum at $q=1$, forming the
soliton centered at $x=x_0$. Therefore  the integral of motion is
given by the potential at the minima: ${\cal E}=\epsilon_0^{(0)}$
and the first term on the r.h.s. of Eq.~(\ref{classaction1}) is
nothing but the equilibrium free energy, Eq.~(\ref{freeshort}).
The activation barrier is then given by the second term on the
r.h.s. of Eq.~(\ref{classaction1}), that is the action of the
soliton:
\begin{equation}\label{actbarrierlong}
U_\infty(\alpha)= U_\infty(0)\, 8\!\!\int\limits_0^{1/2}\!\!
\sqrt{2\left(\epsilon_q^{(0)}(\alpha)
-\epsilon_0^{(0)}(\alpha)\right)}\,\, dq\, ,
\end{equation}
where $U_\infty(0)=eE_0\xi/2$. In the $\alpha\to 0$ limit
$\epsilon_q^{(0)}\to q^2/2$ and Eq.~(\ref{actbarrierlong}) is
trivially satisfied. In general, by analogy with the short channel
case, one may write
$U_\infty(\alpha)=U_\infty(0)f_\infty(\alpha)$, where
$f_\infty(\alpha)=8\int_0^{1/2}\!
\sqrt{2(\epsilon_q^{(0)}-\epsilon_0^{(0)})}\, dq$ is the universal
function of the dimensionless concentration. This function is
plotted in Fig.~\ref{fig4}.

In the low concentration limit $\alpha\ll 1$, employing
Eqs.~(\ref{seconorder}) and (\ref{actbarrierlong}), one arrives at
Eq.~(\ref{weakdopalpha}). As was mentioned above, this expression
is only valid for $1\gg\alpha > \sqrt{x_T/\xi}$. Thus, there is no
true logarithmic singularity in  $\alpha\to 0$ limit.

In the opposite limit, $\alpha >1$, the band $\epsilon^{(0)}_q$ is
harmonic and exponentially narrow:
$\epsilon^{(0)}_q-\epsilon^{(0)}_0=2^{-3}f_0(\alpha)\sin^2\pi q$,
see Eqs.~(\ref{dispersion}), (\ref{bandwidth}). Substituting it
into Eq.~(\ref{actbarrierlong}), one arrives at
Eq.~(\ref{heavydopalpha}). Comparing it with the short channel
result, Eq.~(\ref{heavydop}), one finds that the long channel
limit is applicable when $L> \xi e^{4\sqrt{\alpha}}\gg \xi$. This
reflects the fact that, according to Eq.~(\ref{classaction0}), the
characteristic  size of the soliton grows as an inverse square
root of the potential height. It is worth mentioning that the
applicability criterion of the WKB approximation coincides with
the condition $U_L(\alpha)> k_BT$. Therefore the results, obtained
above, are applicable as long as the activation barrier is
essential.

\subsection{Intermediate length channel}
\label{sec53}

We shall discuss now the channel length's dependence of the
activation barrier. In the limit of  zero concentration, $\alpha
\to 0$, the classical potential takes the form of the periodically
continued  parabola: $\epsilon_q^{(0)}=q^2/2$ for $|q|<1/2$. The
corresponding classical action (\ref{classaction1})  may be easily
found: $S_{L}(q,1-q)=(\xi/2x_T)\tanh (L/2\xi)$, where the optimal
boundary charge is $q=1/(2\cosh(L/2\xi))$. Employing
Eq.~(\ref{freecsp}), one finds Eq.~(\ref{intermediate}) for the
length--dependence of the activation barrier. It may be easily
derived  considering  the interaction energy,
Eq.~(\ref{totalenergy}), with the interaction  potential,
Eq.~(\ref{modelpotential}), of  three charges: $q$ and $1-q$ at
the two boundaries and $-1$ in the middle. Optimizing the energy
with respect to $q$, one obtains Eq.~(\ref{intermediate}).

Yet another derivation of Eq.~(\ref{intermediate}) may start from
considering two membrane -- water interfaces at $x =\pm L/2$ as
metallic planes creating the series of negative images at $|x| =
(2k-1)L$ and positive ones at $|x| = 2kL$, where $k$ is a positive
integer. All the images are located along the channel axis. The
self--energy of the central charge is $U_{L} = e\phi(0)/2$, where
$\phi(0)$ is the potential of all  charges at $x=0$. It consists
of the potential of the central charge, $\Phi(0)$, and the sum of
potentials off all the images, $\phi_i(0)$:
\begin{equation}
\phi_{i}(0)= 2E_{0}\xi(0)\sum\limits_{k =1}^{\infty}
(-1)^{k}e^{-kL/\xi}= \frac{-2E_{0}\xi}{e^{L/\xi}+1}\, ,
\label{images}
\end{equation}
where we have used Eq.~(\ref{modelpotential}). Adding up
$\phi_{i}(0)$ and $\Phi(0)$ and taking into account that
$U_{\infty}(0) = 2eE_{0}\xi$ (c.f. Eq.~(\ref{long})),  one arrives
at Eq.~(\ref{intermediate}).

For not very long channel the string  does not go far away from
the maximum of the potential: $\epsilon^{(0)}_{1/2}$. Thus one may
expand the potential to the second order around the maximum:
$\epsilon^{(0)}_q\approx \epsilon^{(0)}_{1/2} -\omega_b^2 \delta
q^2/2$, where $\omega_b^2 \equiv -\partial^2_q
\epsilon^{(0)}_q|_{q=1/2}$ and $\delta q= q-1/2$. Substituting
this form of the potential into the definition of the canonical
momentum, $p(q)$, and then  in  Eq.~(\ref{classaction1}), one
finds for the saddle point action:
\begin{equation}\label{criticalactiaon}
  S^{sp}_L= {2 \epsilon^{(0)}_{1/2}L\over x_T} - {2\delta{\cal
  E}\over x_T} \left(L - {\pi\xi \over \omega_b}  \right)\, ,
\end{equation}
where the integration ran over $|\delta q|\leq \sqrt{2\delta{\cal
E}}/\omega_b$ and $\delta{\cal E}\equiv  \epsilon^{(0)}_{1/2}
-{\cal E}
 >0$ may be found using the saddle point approximation. As a
result, for $L\leq L_c(\alpha)$, where
\begin{equation}\label{criticallength}
  L_c(\alpha)={\pi\xi \over \omega_b}=\pi\xi \left( -\partial^2_q
\epsilon^{(0)}_q|_{q=1/2}\right)^{-1/2}\, ,
\end{equation}
the optimal value is $\delta{\cal E}=0$ and the saddle point
action is $S^{sp}_L= 2 \epsilon^{(0)}_{1/2}L/ x_T$. This is
exactly the same result as without electric field escape from the
channel, cf. Eq.~(\ref{Zq1-q}). Therefore, deviations of the
potential from the ideal 1d Coulomb law are relevant only for $L>
L_c$.

Employing Eqs.~(\ref{point1/2}) and (\ref{dispersion}), one finds
for the threshold length $L_c(\alpha) = \pi\sqrt{2\alpha}\,\xi$
for $\alpha\ll 1$ and $L_c(\alpha)=2\xi/\sqrt{f(\alpha)}\sim \xi
e^{4\sqrt{\alpha}}$ for $\alpha > 1$. In agreement with
Eq.~(\ref{intermediate}), $L_c(0)=0$ and the crossover from the
short to the long channel is smooth at $\alpha\to 0$. On the
contrary, at a finite concentration of salt the length dependence
of the activation barrier exhibits non-analyticity at $L=L_c$. For
$L\leq L_c$ the activation barrier is strictly linear in $L$,
Eq.~(\ref{actbarrier1}), while it smoothly saturates to
Eq.~(\ref{actbarrierlong}) at $L >L_c$. The observables, such as
the resistance, do not exhibit  non-analyticity due to the
pre-exponential factors. The very similar phenomena takes place in
the crossover from the thermally  activated escape from a
potential well  to the quantum tunnelling \cite{Afleck81}.

\section{Narrow Channel}
\label{sec6}

The previous consideration was based on the model of a pure 1d
motion. The latter is justified if the characteristic size of the
thermal pairs is significantly larger than the channel radius:
$x_T\gg a$. Recalling that $x_T=a ^2/(2l_B)$, one finds that the
applicability condition of the 1d treatment is: $l_B\ll a \ll L$.
These strict inequalities are rarely satisfied for real ion
channels. The purpose of this section is to analyze how the
opposite case $a\lesssim l_B$ may be incorporated into the theory
presented above. The latter condition may be formulated as
$k_BT\lesssim e^2/(\kappa_1 a)$, meaning that the interaction
energy on the scale $a$ is statistically significant.

To take into account the transverse degrees of freedom, one may
repeat the calculations of section \ref{sec4}, introducing the 3d
density, $\rho(\vec r)$, and conjugated field, $\theta(\vec r)$.
The action for the latter takes the form (c.f.
Eq.~(\ref{greenfunct})):
\begin{equation}\label{3daction}
  S[\theta]=\int\!\! d^3 \vec r\left[{1\over 8\pi l_B}(\nabla_{\vec
r}\,\theta)^2 -
  2c\cos\theta\right]  \, ,
\end{equation}
where the 3d integral runs over the volume of the channel. The
potential field, $\theta(\vec r)$, may be parameterized as:
$\theta(\vec r)=\theta_0(x)+\delta\theta(x,\vec\varrho)$, where
$\vec\varrho$ stands for the two transverse coordinates. The
non--uniform part, $\delta \theta(x,\vec\varrho)$, may be expanded
in the basis of the transverse modes of the Laplace operator:
$\nabla_{\vec\varrho}^2\varphi_s = -\lambda_s^2 \varphi_s$, where
$\varphi_s(\vec\varrho)$ have zero normal derivative at the
channel boundary, and we use normalization $\int d^2\vec\varrho \,
\varphi_s^2 = \pi a^{\,2}$. As a result:
\begin{equation}\label{laplaceexpansion}
  \theta(x,\vec\varrho)=\theta_0(x) +\sum\limits_{s}
  \theta_s(x)\, \varphi_s(\vec\varrho)\, .
\end{equation}
In terms of the new set of fields, $\theta_s(x)$, the gradient
term of the action (\ref{3daction}) takes the form:
\begin{equation}\label{S0}
  S_0[\theta]={x_T\over 4}\!\!\! \int\limits_{-L/2}^{L/2}\!\!\! d
x\left[(\partial_x\theta_0)^2
  + \sum\limits_{s} \left((\partial_x\theta_s)^2
  +\lambda_s^2 \theta_s^2\right)\right]  \, .
\end{equation}
It is evident from this expression that $\langle
\theta_s^2\rangle=(x_T\lambda_s)^{-1}\sim l_B/a$. Therefore in the
limit $a\gg l_B$, the transverse fluctuations may be neglected,
$\delta\theta\ll 1$, and the theory reduces to that of the single
field, $\theta_0(x)$, in agreement with the previous sections. In
the opposite limit, $a\lesssim l_B$, one may average the cosine
term on the r.h.s. of Eq.~(\ref{3daction}) over the fluctuations
of $\theta_s(x)$ which are governed by the action (\ref{S0}). As a
result, one  obtains the effective action of the field
$\theta_0(x)$ of the form of Eq.~(\ref{greenfunct}) with the
renormalized amplitude of the cosine term:
\begin{equation}\label{alfaeff}
  \alpha_{eff}\equiv cx_T\!\!\int\!\! d^2\vec\varrho\left\langle
  e^{i\delta\theta}\right\rangle_{\theta_s} = 2\pi
cx_T\!\!\!\int\limits_0^{a-b}\!\!
  d\varrho\,\varrho\, e^{-U_{eff}(\varrho)/(k_BT)}\,  ,
\end{equation}
where the upper limit of the radial integration is limited  by the
radius of the ion, $b$. The effective potential in the transversal
direction is given by
\begin{equation}\label{Ueffective}
  U_{eff}(\varrho)={e^2\over \kappa_1 a^{\,2}}\lim_{x\to 0} \left[
  \sum\limits_s\frac{\varphi_s^2(\varrho)\,e^{-\lambda_s|x|}}{\lambda_s}
  - {a^{\,2}\over 2|x|}\right]\, .
\end{equation}
We have subtracted the self--energy of the ion in the bulk
solution: $e^2/(2\kappa_1|x|)$ to regularize the expression at
small distances, $|x|\ll a$. Indeed, such a local contribution to
the self--energy is not affected by moving the ion from the bulk
into the channel. Thus it must be subtracted  from the chemical
potential, $\mu$, to have the proper definition of the bulk
concentration, $c$.

\begin{figure}[ht]
\begin{center}
\includegraphics[height=0.18\textheight]{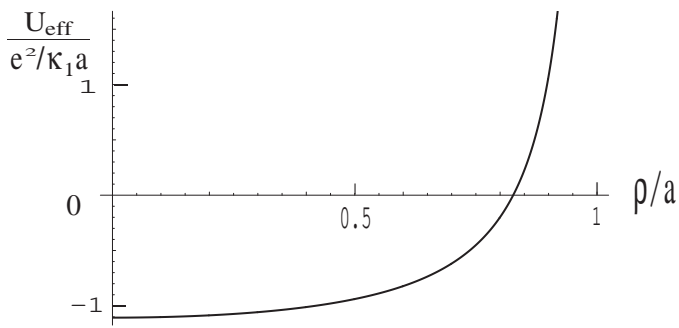}
\end{center}
\caption{Effective potential in units of $e^2/(\kappa_1 a)$ as a
function of the distance from the axis of the cylindrical
channel.} \label{fig6}
\end{figure}

For the cylindrical channel, the effective potential may be
expressed in the closed form through the modified Bessel
functions~\cite{Smythe}:
\begin{equation}\label{Ueffective1}
  U_{eff}(\varrho)={e^2\over \kappa_1 a}
 \int\limits_0^\infty\! {dk\over \pi}\left[ -{2\over k^2}-
 \!\!\sum\limits_{n=-\infty}^\infty\!\!
 I_n^2\left(k{\varrho\over a}\right)
 \frac{K_n'(k)}{I_n'(k)}\right]  \, .
\end{equation}
The integral on the r.h.s. of this expression is plotted in
Fig.~\ref{fig6} versus $\varrho/a$. It diverges as
$a/[4(a-\varrho)]$ near the channel's boundary, indicating
repulsion of the ion from the charge image in the media with
smaller dielectric constant. This divergence plays a minor role in
Eq.~(\ref{alfaeff}), however, because  $a-\varrho \geq b$ there.

Remarkably, the effective potential is {\em negative} in the
middle of the channel: $U_{eff}(0)\simeq -1.1e^2/(\kappa_1 a)$.
This means that the ions have a preference to go inside the
channel, provided that the long--range part of the self--energy,
$\propto L$, is not included. As before, the long--range part of
the interaction energy is described by the field $\theta_0$, which
is uniform in the transversal direction. This  long--range part of
the ion's self--energy  is related to the uniform electric field
$E_0$ at distance $|x|\gg a$ from the charge. At smaller distance,
$|x|\lesssim a$ the field is almost spherically symmetric, the
same as in the bulk. As a result, the near vicinity of the charge,
$|x|\lesssim a$, does not contribute to the excess self--energy.
This ``saves'' the energy $1.1e^2/(\kappa_1 a)$ per ion.

Let us return to the  collective saddle point  with the
half--integer boundary charges, discussed in section \ref{sec32}.
Such boundary charges produce the electric field $E_0$, creating
the OFP state. When an additional ion from the bulk enters the OFP
state of the channel, its long range part of the self--energy is
cancelled (by the field $E_0$ of the boundary charges), but the
energy gain $1.1e^2/(\kappa_1 a)$ remains. One may say  that each
charge reduces the length of the string of the field $E_0$
approximately by $a$, ``saving'' the energy $1.1e^2/(\kappa_1 a)$.
This energy gain comes on top of the entropy gain, discussed in
section \ref{sec32}. It leads to the increase of the effective
concentration $\alpha_{eff}$, Eq.~(\ref{alfaeff}), in comparison
with the naive geometrical one, $\alpha=cx_T\pi a^{\,2}$.

\begin{figure}[ht]
\begin{center}
\includegraphics[height=0.2\textheight]{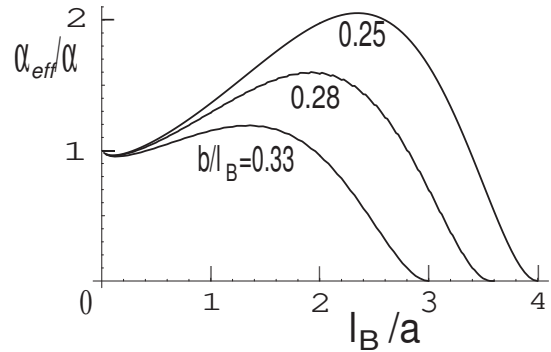}
\end{center}
\caption{$\alpha_{eff}/\alpha$ as a function of $l_B/a$ for the
three values of $b/l_B$.} \label{fig7}
\end{figure}

The ratio $\alpha_{eff}/\alpha$ for three choices of the hydrated
ion radius $b$ is plotted versus $l_B/a$ in Fig.~\ref{fig7}. One
notices that for a reasonable for NaCl and KCl
value~\cite{Robinson} $b=0.28 l_B=0.2$ nm (used in section
\ref{sec8} for numerical estimates) the effective concentration
may be up to $1.6$ times larger than the geometric one (the latter
does {\em not} take into account the finite core radius). The
results of the previous sections, summarized in section
\ref{sec2}, may be straightforwardly applied now with the
$\alpha_{eff}$ substituted for $\alpha$.

 \section{Dynamics}
\label{sec7}


Having discussed at length the activation barrier and its
concentration and geometry dependence, we turn now to the
pre-exponential factor in the expression for the channel
resistance, Eq.~(\ref{conductance}).  To this end we need a
dynamical model for  the ion's motion inside the channel. It will
also allow us to derive non--linear voltage--current relations,
discussed in section \ref{sec35}. We shall assume that the ions
undergo dissipative brownian motion due to viscous friction with
the water. The latter eventually transfers an excess momentum, the
ions acquire in the external electric field, to the walls of the
channel.  Under these assumptions one may write the Langevin
equation for the time--dependent coordinate of the $i$-th ion,
$x_i(t)$:
\begin{equation}
\gamma \dot x_i = \sigma_i {V\over L} -{\partial U\over \partial
x_i} +\eta_i(t)\, ,
                                         \label{xlangevin}
\end{equation}
where $V$ is an externally applied voltage, $\gamma$ is the
viscous friction coefficient and $\eta_i(t)$ is the Gaussian noise
with the amplitude given by the fluctuation--dissipation theorem:
\begin{equation}\label{langevinnoise}
  \langle \eta_i(t) \eta_j(t') \rangle = 2\gamma k_BT
  \delta_{ij}\delta(t-t')\, .
\end{equation}
In the absence of the activation barrier and the interparticle
interactions these assumptions lead to the channel resistance
given by the Drude formula: $R_0\equiv \gamma L/ (2n)$. For a very
small concentration $\alpha\to 0$, the ions cross the channel one
by one overcoming the activation barrier given by
Eq.~(\ref{parabolic}) (but not interacting with other ions). The
resistance associated with such a process is:
\begin{equation}\label{R00}
  R=R_0\sqrt{\pi x_T\over L}\,\, e^{U_L(0)/(k_BT)}\,.
\end{equation}

We shall  assume furthermore that the resistance of the bulk
reservoirs is negligible in comparison with that of the channel.
This means that the boundary charges adjust  to an instantaneous
configuration of ions inside the channel. The boundary charge
$q(t)$ is thus found from the minimum of the interaction energy:
$\partial U(q,1-q)/\partial q = 0$. We have assumed here that
there is a single negative excess charge, travelling across the
channel,  and thus $q+q'=1$.  From Eq.~(\ref{totalenergy_q}) one
then obtains:
\begin{equation}\label{qoft1}
    q(t)={1\over 2}+{1\over L}\sum\limits_i \sigma_i x_i(t)\, .
\end{equation}
This relation imposes  the constraint on possible configurations
of ions, $x_i$.  Such a constraint  was not included in the
thermodynamic calculations. To discuss the dynamics one needs to
define the equilibrium {\em constrained} free energy as:
\begin{equation}\label{constrZ}
 e^{-\frac{F^*_L(q)}{k_BT}} \equiv \left\langle \delta\left({1\over
L}\sum\limits_i \sigma_i
  x_i(t) + {1\over 2} - q \right) \right\rangle_{x_i}  ,
\end{equation}
where the angular brackets denote summation over number of ions
and integrations over their coordinates with the weight given by
Eq.~(\ref{partition}). Introducing the Lagrange multiplier, one
obtains the relation between the constrained  and the
unconstrained free energies:
\begin{equation}\label{Fstar}
e^{-\frac{F^*_L(q)}{k_BT}}={iL\over \pi x_T}\!\!\!
\int\limits_{1/2-i\infty}^{1/2+i\infty}\!\!\!\!\! dQ\,\,
e^{\,-\frac{F(Q)}{k_BT} + (Q-q)^2L/x_T} .
\end{equation}
This integral  may be evaluated in the saddle point approximation.
From the saddle point equation one  observes that the {\em
extensive} parts of the constrained and the unconstrained free
energies coincide at the two extrema, $q=0$ and $q=1/2$. This is
the reason why one can discuss the activation barrier using the
unconstrained free energy. The two free energies  have  distinct
non-extensive parts, however, which must be kept to determine the
pre-exponential factor.

We multiply now the constraint, Eq.~(\ref{qoft1}), by  the Drude
resistance, $R_0=\gamma L/(2n)$, differentiate it over time and
employ the Langevin equations (\ref{xlangevin}). As a result:
\begin{equation}\label{dynamic}
R_0 \dot q(t) = V -{1\over 2n}\sum\limits_i\sigma_i {\partial
U\over
\partial x_i} +\eta(t)    \, ,
\end{equation}
where $\eta(t)\equiv (2n)^{-1}\sum_i \sigma_i\eta_i(t)$. This is
the Gaussian noise with the correlator:
\begin{equation}\label{noiseR}
 \langle \eta(t) \eta(t') \rangle =\sum\limits_{i,j}
 {\sigma_i \sigma_j\over (2 n)^2}
 \langle \eta_i(t) \eta_j(t') \rangle = 2k_BT R_0\, \delta(t-t')\, .
\end{equation}
It is thus the equilibrium Nyquist  noise of the Drude resistor
$R_0$.

Since the boundary charge, $q(t)$, is a weighted average of many
uncorrelated fast functions, $x_i(t)$, one may assume that it
possesses  a slow dynamics. Thus, one may average out the fast
motion of individual ions, $x_i(t)$, over the equilibrium
distribution obtained for an instantaneous adiabatic value of
$q(t)$. Averaging this way the second term on the r.h.s. of
Eq.~(\ref{dynamic}), one obtains \cite{footnote}

\begin{equation}\label{final}
  R_0 \dot q = V- {\partial F^*_L(q)\over \partial q} +
  \eta(t)\, .
\end{equation}
The corresponding Fokker--Planck  equation for the probability
distribution function $P(q,t)$ is:
\begin{equation}\label{FP1}
  \partial_t P = -\partial_q I\,; \hskip .5cm
  I\equiv {1\over R_0} \left[P\left(V-\partial_q
  F_L^*\right)-k_BT\partial_q
  P\right]\, .
\end{equation}
Following the standard route \cite{Ambegaokar69}, one obtains for
the linear resistance, $R=V/I$, in the stationary case, when the
current $I$ is a constant:
\begin{equation}\label{R0}
R = R_0 \int\limits_0^1\!\! dq\, e^{F_L^*(q)/(k_BT)}
\int\limits_0^1\!\! dq\, e^{- F_L^*(q)/(k_BT)}\, .
\end{equation}
The first  integral here is dominated by the vicinity of the
barrier top, $q=1/2$, while the second one -- by the equilibrium
state $q=0,1$. As a result, the resistance is given by:
\begin{equation}\label{R}
R =  R_0\, {\cal A}_L(\alpha)\, e^{\, U_L(\alpha)/(k_BT)}\, ,
\end{equation}
where $U_L(\alpha)=F_L(1/2)-F_L(0)$ is the activation barrier and
factor ${\cal A}_L(\alpha)$ originates from the fluctuations
around the extremal   values of the integrals in
Eqs.~(\ref{Fstar}) and (\ref{R0}).  Calculating the Gaussian
integrals, one obtains for the pre-exponential factor:
\begin{equation}\label{A}
\!\!{\cal A}_L(\alpha) = {\pi x_T\over L}\,
\frac{1+\omega_b^2}{\omega_0\omega_b} \, .
\end{equation}
where $\omega_0^2 \equiv \partial^2_q \epsilon^{(0)}_q|_{q=0}$ and
$\omega_b^2 \equiv -\partial^2_q \epsilon^{(0)}_q|_{q=1/2}$.

In the dilute limit, $\alpha\ll 1$,  from Eqs.~(\ref{seconorder})
and (\ref{point1/2}) one obtains: $\omega_0\approx 1-4\alpha^2$
and $\omega_b\approx (2\alpha)^{-1/2}$ and thus ${\cal
A}_L(\alpha) \approx  \pi x_T /( L \sqrt{2\alpha})$. The
divergence at small $\alpha$ is terminated at $\alpha\sim x_T/
L\ll 1$ (that is $nL\sim 1$, so there is less than one particle in
the channel in average). At such a small concentration one can not
assume that the $q(t)$ is a collective slow degree of freedom.
Formally, the integrals in Eq.~(\ref{R0}) can not be considered as
Gaussian and the pre-factor saturates at $ {\cal A}_L(0)
=\sqrt{\pi x_T/ L}$ as required by Eq.~(\ref{R00}).

In the dense limit, $\alpha > 1$, one finds that $ F^*(q)\approx
F(q)$ and thus  $\omega_0=\omega_b= \pi \sqrt{f_0(\alpha)}/2\ll
1$. As a result, ${\cal A}_L(\alpha) = 4x_T /(\pi L
f_0(\alpha))=\pi^{-1}k_BT/U_L(\alpha)$. The pre--factor grows with
the concentration. This growth may be associated with the slow
motion of the collective variable $q(t)$ near the top of the
activation barrier.  This result is applicable as long as
$U_L(\alpha)/(k_BT) > 1$.

To conclude, we found that the pre--exponential factor ${\cal
A}(\alpha)$, cf. Eq.~(\ref{R}),  exhibits the {\em
non--monotonous} dependence on the salt concentration. It reaches
a flat minimum at $\alpha \approx 0.2$, where ${\cal
A}_L(.2)\approx 8x_T/L$ and increases  both for smaller and larger
concentrations. Employing Eq.~(\ref{FP1}), one may go beyond the
linear response theory and discuss non--linear voltage--current
relation. Since such  a derivation is relatively standard
\cite{Ambegaokar69} we do not provide it here. The corresponding
results are listed in sections \ref{sec1} and \ref{sec35}.

\section{Conclusions}
\label{sec8} In this paper we suggested a theory of the
salt--induced reduction of the transport activation barrier in the
water--filled channels in a media with much smaller dielectric
constant, than that of the water. Our theory is based on
one--dimensional nature of the electric field in the channel. We
have derived a universal dependencies of the barrier height
$U_{L}(\alpha)$ as a function of the dimensionless parameter
$\alpha$, given by Eq.~(\ref{alpha}). They are presented in
section \ref{sec2}. Remarkably, when the channel radius $a$ grows,
the one--dimensional character of the problem remains valid until
at some $\alpha$ the barrier $U_{L}(\alpha)$ becomes smaller than
$k_{B}T$. In this sense one may say that transition from
one-dimensional channel to the bulk is governed solely by $\alpha$
(and {\em not} any other parameter, e.g. $\xi/a$, or $x_D/a$,
etc.)

Now we would like to consider  a few examples of application of
our theory to water--filled pores. The first one is
$\alpha$-hemolysin, a passive protein channel in  lipid membranes.
As we mentioned above, for the water--filled channels in lipid
membranes $\xi\simeq 6.8\, a$. For the $\alpha$-hemolysin $L= 5$
nm, $a = 0.75$ nm and thus $L/\xi \approx 1$. For this case,
according to Eq.~(\ref{long}) we obtain $U_{\infty}(0) = 6.8
k_{B}T_{r}$ and according to Eq.~(\ref{intermediate}) $U_{L}(0) =
0.46U_{\infty}(0) = 3.1 k_{B}T_{r}$, where $T_r$ is the room
temperature. To describe the effect of salt one can use the curve
for $L/\xi = 1$ in Fig.~\ref{fig4} with $\alpha_{eff}$ substituted
for $\alpha$. The point where $U_{L}(\alpha)/U_{L}(0) = 0.5$,
corresponds to $\alpha_{eff} = 0.2$. For  $a = 0.75$ nm, $l_B/a =
0.93$ Fig.~\ref{fig7} gives $\alpha = 0.8\alpha_{eff}= 0.16$,
corresponding to $c = 0.37$ M and the bulk concentration of salt
$c_0 = 0.46$ M. Thus, it is enough to have $0.5$ M of salt to
reduce the activation exponent by the factor of two.

The second example is a water--filled nanopore made in a silicon
film ~\cite{Li,Storm}. Such nanopores are considered in
nanoscience for the transport of DNA, in salty water~\cite{Timp}.
We are concerned only with the transport of small ions in a e.g.
KCl solution. Let us assume that $a = 1$ nm and the length $L =
20$ nm. Silicon oxidizes around the channel, giving $\kappa_2
\simeq 4$ and $\kappa_1/\kappa_2=20 \gg 1$. Using
Eq.~(\ref{criticaleq}), we find $\xi=4.75\, a$. This gives $L/\xi
\simeq 4$. For this case, according to Eq.~(\ref{long}) we obtain
$U_{\infty}(0) = 3.6 k_{B}T_{r}$ and Eq.~(\ref{intermediate})
results in $U_{L}(0) = 0.96 U_{\infty}(0) = 3.4 k_{B}T_{r}$.
Effect of salt is given by the top full curve in Fig.~\ref{fig4},
corresponding to $L/\xi = 4$. In this case,
$U_{L}(\alpha)/U_{L}(0) = 0.5$ at $\alpha_{eff} = 0.4$ or at
$\alpha = 0.36$ (as we see in Fig.~\ref{fig7} for $l_B/a =0.7$ the
ratio $\alpha_{eff}/\alpha = 1.12$). This requires $c = 0.26$ M
and $c_{0} = 0.31$ M.

The third example is a water--filled nanopore with $L = 40$ nm and
$a=1.5$ nm in cellulose acetate film used for inverse osmosis
desalination. Let us take $\kappa_2 \simeq 2$ so that $\xi \simeq
6.8\, a \simeq 10$ nm , $L/\xi \simeq 4$, $U_{L}(0) = 0.96
U_{\infty}(0) \approx 3 k_{B}T_{r}$.  Effect of salt is again
given by the top full curve in Fig.~\ref{fig4} and
$U_{L}(\alpha)/U_{L}(0) = 0.5$ at $\alpha_{eff} = 0.4$ which in
this case corresponds to $\alpha =0.38$. This happens at $c =
0.69$ M and $c_{0} = 0.73$ M.

The last example, we want to consider, is the gramicidin-A-like
channel in a lipid membrane. In this case we use $L= 2.5$ nm, $a =
0.3$ nm and obtain $L/\xi =1.2$. According to Eq.~(\ref{long}) the
barrier $U_{\infty}(0) = 17 k_{B}T_{r}$ and according to
Eq.~(\ref{intermediate}): $U_{L}(0) = 0.55\, U_{\infty}(0) = 9.4
k_{B}T_{r}$. We see that, because of the very small width, we deal
with much higher barriers. On the other hand, at such a small $a$
even the concentration of salt $c_0$ close to the saturation limit
of  $6$ M and corresponding $c=5.2$ M we can reach only $\alpha =
0.06$ or $\alpha_{eff} = 0.084$ (for this case $l_{B}/a = 2.3$ and
according to Fig.~\ref{fig7} $\alpha_{eff}/\alpha = 1.4$). At this
$\alpha_{eff}$ interpolating between curves for $L/\xi = 1$ and
$L/\xi = 1.5$ we arrive at $U_{L}(\alpha)=0.83 U_{L}(0)= 7.8
k_{B}T_{r}$. Thus, the effect of salt  on a narrow channel is
 weak. This could be anticipated from
Eq.~(\ref{alpha}), if we rewrite it as $\alpha = \pi
(cl_{B})^{3}(a/l_B)^{4}$ and take into account that
$cl_{B}^{3}\leq 1$, even at the saturation limit $c_{0}=6$ M.
Contrary to the case where $a/l_{B}\geq 1$ and the range $ \alpha
> 1$ is accessible, in the opposite limit $a/l_{B}\lesssim  1$ the
parameter $\alpha$ is bound to be small~\cite{footnote2}.

In this paper we considered only channels without charges fixed on
their walls. Methods used here can be easily applied for channels
with fixed charges of one or both signs. We will address such
channels in the next publication.


\section{Acknowlegments}
\label{sec9}

We are grateful to A. L. Efros, M. M. Fogler, L.~I.~Glazman, A.
Yu. Grosberg, A. Meller, S. Teber, Ch. Tian and M. Voloshin for
numerous useful discussions. A.~K. is supported by the A.P. Sloan
foundation and the NSF grant DMR--0405212. A.~I.~L. is supported
by NSF Grants No. DMR-0120702 and DMR-0439026. B.~I.~S is
supported by NSG grant DMI-0210844.



\begin{thebibliography}{99}
\bibitem{footnote} To carry out this averaging one parameterizes the ions
 coordinates as $ x_i(t)= \sigma_i (q(t)-1/2)/(2
n) +\delta x_i(t)$, where the constraint (\ref{qoft1}) is reduced
to $\sum_i \sigma_i \delta x_i(t) =0$. The second term on the
r.h.s. of Eq.~(\ref{dynamic}) may be written now as:
$$ \,\,\,\, \left. {\partial U\over \partial q}\right|_{\{\delta
x_i\}}\!\!\!\!\! =\!\!
 \left. {\partial U\over \partial q}\right|_{\{x_i\}}\!\!\!\!\!
 +\!
\sum\limits_i\left. {\partial U\over \partial x_i}\right|_q \left.
{\partial x_i\over \partial q}\right|_{\{\delta
 x_i\}}\!\!\!\!= {1\over 2n} \sum\limits_i \sigma_i\left. {\partial U\over
\partial
 x_i}\right|_q\, ,$$
where we employed that $\partial U/\partial q|_{\{x_i\}}=0$.
Averaging over  the equilibrium distribution function
$Z^{-1}\exp\{-U(\{\delta x_i\},q(t))/(k_BT)\}$, one  finds:
$(2n)^{-1}\langle \sum_i \sigma_i
\partial U/ \partial x_i|_q \rangle =\langle \partial U/
\partial q|_{\{\delta x_i\}} \rangle = \partial F^*_L(q)/\partial
q$.
\bibitem{footnote2} Another possible application of our theory is
the transport in
single-crystal silicon nanowires~\cite{Lieber} suspended in air
between two conductors. They have $ a \sim 1$ nm and lengths up to
several microns. In this case, $\kappa_1 = 11$ and $\kappa_2 =1 $
so that the ratio $\kappa_{1} /\kappa_2$ is  relatively large. We
will address this application in a future publication.
\end{thebibliography}
\end{document}